\documentclass[aps,prd,amsfonts,onecolumn,preprintnumbers,superscriptaddress,showpacs,floatfix,nofootinbib]{revtex4}

\usepackage{graphicx}
\usepackage[T1]{fontenc}
\usepackage[latin9]{inputenc}

\newcommand{\be}{\begin{equation}}
\newcommand{\ee}{\end{equation}}
\newcommand{\bea}{\begin{eqnarray}}
\newcommand{\eea}{\end{eqnarray}}

\newcommand{\gapp}{\mathrel{\raise.3ex\hbox{$>$}\mkern-14mu
              \lower0.6ex\hbox{$\sim$}}}
\newcommand{\lapp}{\mathrel{\raise.3ex\hbox{$<$}\mkern-14mu
              \lower0.6ex\hbox{$\sim$}}}

\newcommand\lsim{\lesssim}
\newcommand\gsim{\gtrsim}

\newcommand\vev[1]{{\langle {#1} \rangle}}
\renewcommand\({\left(}
\renewcommand\){\right)}
\renewcommand\[{\left[}
\renewcommand\]{\right]}

\newcommand\eq[1]{Eq.~(\ref{#1})}
\newcommand\eqs[2]{Eqs.~(\ref{#1}) and (\ref{#2})}
\newcommand\eqss[3]{Eqs.~(\ref{#1}), (\ref{#2}), and (\ref{#3})}

\newcommand\eqreff[1]{(\ref{#1})}
\newcommand\eqsref[2]{(\ref{#1}) and (\ref{#2})}

\newcommand\pa{\partial}

\newcommand\mpl{M_{\rm P}}

\def\cala{{\cal A}}
\def\calb{{\cal B}}
\def\cald{{\cal D}}

\def\call{{\cal L}}

\def\caln{{\cal N}}
\def\calp{{\cal P}}

\def\calv{{\cal V}}
\def\calpz{{\calp_\zeta}}

\newcommand\bfA{{\mathbf A}}

\newcommand\bfd{{\mathbf d}}

\newcommand\bfk{{\mathbf k}}

\newcommand\bfn{{\mathbf N}}

\newcommand\bfp{{\mathbf p}}

\newcommand\bfx{{\mathbf x}}

\newcommand\GeV{\,\mbox{GeV}}
\newcommand\MeV{\,\mbox{MeV}}

\newcommand\sub[1]{_{\rm #1}}
\newcommand\su[1]{^{\rm #1}}

\newcommand\mone{^{-1}}
\newcommand\mtwo{^{-2}}
\newcommand\mthree{^{-3}}

\newcommand\mhalf{^{-1/2}}
\newcommand\half{^{1/2}}

\newcommand\mthreehalf{^{-3/2}}
\newcommand\mn{{\mu\nu}}

\newcommand{\fnl}{f\sub{NL}}

\begin{document}


\title{Statistical anisotropy of the curvature perturbation from vector field perturbations}

%\title{Vector field perturbations and 
%statistical anisotropy of the curvature perturbation}
\author{Konstantinos Dimopoulos}
\email{konst.dimopoulos@lancaster.ac.uk}
\affiliation{Department of Physics, Lancaster University, Lancaster LA1 4YB, UK\
}
\author{Mindaugas Kar\v{c}iauskas}
\email{m.karciauskas@lancaster.ac.uk}
\affiliation{Department of Physics, Lancaster University, Lancaster LA1 4YB, UK\
}
\author{David H. Lyth}
\email{d.lyth@lancaster.ac.uk}
\affiliation{Department of Physics, Lancaster University, Lancaster LA1 4YB, UK\
}
\author{Yeinzon Rodr\'{\i}guez}
\email{yeinzon.rodriguez@uan.edu.co}
\affiliation{Centro de Investigaciones, Universidad Antonio Nari\~no, Cra 3 Este \# 47A-15, Bogot\'a D.C., Colombia}
\affiliation{Escuela de F\'{\i}sica, Universidad Industrial de Santander, Ciudad Universitaria, Bucaramanga, Colombia}

\preprint{PI/UAN-2008-300FT}

\begin{abstract}

The $\delta N$ formula for 
the primordial curvature perturbation $\zeta$ 
is extended  to include vector  as 
well as scalar fields. Formulas for the tree-level contributions to the 
spectrum and bispectrum of $\zeta$ are given, exhibiting statistical
anisotropy. The one-loop contribution to 
the spectrum of $\zeta$ is also worked out.  We then consider the
 generation of  vector field perturbations  from the vacuum, including
the  %previously ignored 
 longitudinal component that will  be present
if there is no gauge invariance.
Finally, the  $\delta N$ formula is  applied
to the vector curvaton and vector inflation models with the tensor perturbation
also evaluated in the latter case. 
\end{abstract}

\maketitle

\section{Introduction}

\label{sintro}

 Starting at an `initial'
temperature of a few MeV, the observable Universe is now
understood in considerable detail. 
At the initial epoch the expanding Universe is
an almost isotropic and homogeneous gas. The perturbations away from
perfect isotropy and homogeneity 
at the initial epoch
are the subject of intense study at present, because they determine
the subsequent evolution of all cosmological perturbations \cite{abook}.
According to observation, the dominant and perhaps the only
initial perturbation is
the curvature perturbation $\zeta$, so-called because it is related
to the perturbation in the intrinsic curvature of space-time slices with
uniform energy density.

To understand the nature and origin of $\zeta$, 
one 
uses comoving coordinates $\bfx$, that move with expansion of the unperturbed
Universe.  Also, one considers
 the Fourier components
with comoving wave-vector $\bfk$. Physical positions are $a(t)\bfx$ and
physical wave-vectors are ${\bf k}/a(t)$, where $a$ is the scale factor of the
Universe. The Hubble parameter is $H \equiv \dot a/a$, with a dot denoting
derivative with respect to the cosmic time $t$. 

It is convenient to smooth all relevant quantities on a comoving scale, 
somewhat
below the shortest scale of cosmological interest. 
This will not affect the Fourier components on cosmological scales,
 and will greatly simplify the analysis. Consider a given cosmological
scale, characterised by wavenumber $k/a$.
On the assumption
that  gravity slows down the expansion of the
 cosmic fluid,  $aH/k = \dot a/k$ increases as we go back in time.
At the present epoch scales of cosmological interest correspond to
$10^{-6}\lsim aH/k\lsim 1$, but at the `initial' temperature $T\sim \MeV$
they all correspond to $aH/k\gg 1$. Such scales are said to be outside
the horizon. 

To explain the origin of the perturbations,
it is supposed that going further back in time we reach an era
of inflation  when by definition gravity is repulsive. At the begining of 
inflation the smoothing scale is supposed to be inside the horizon.
With mild assumptions, it can be shown that inflation drives all perturbations
to zero at the classical level.  But as each scale $k$ leaves
the horizon, the quantum fluctuations  of those  scalar field perturbations
  with
mass $m\lsim H$ are  converted  \cite{ls,class} to classical perturbations.

According to the usual assumption, one  or more of these scalar field
perturbations is responsible for the curvature perturbation 
(for a recent account with references see  Ref. \cite{mycurv}).
In that case, the statistical properties of $\zeta$ (specified by its
correlators) are homogeneous and isotropic (invariant under displacements
and rotations). 
It has been pointed out recently that vector field perturbations
 could contribute to 
$\zeta$ \cite{VC,supergvc,RA2,ys}\footnote{Non-standard spinors may be used for the same purpose. See Ref. \cite{bohmer}.}.
Such contributions will typically make $\zeta$
statistically anisotropic, but still statistically homogeneous.

It was shown  in an earlier paper \cite{DY} how, 
including only scalar fields, one may calculate
the correlators of $\zeta$ through what is called the $\delta N$ formalism 
\cite{starobinsky,ss,lms}.
The $\delta N$ formalism has recently been applied to the vector
field case in a particular setup \cite{ys}. In this paper, we work out
a completely general $\delta N$ formalism including vector fields
and then apply it to a different setup used for the
vector curvaton \cite{VC,supergvc,RA2} and vector inflation \cite{VI} 
 scenarios.

The plan of the paper is the following.  
In Section \ref{sobs} we give some useful formulas and survey the observational 
status regarding statistical anisotropy.  Section \ref{sdeln} is devoted to a brief 
description of the $\delta N$ formalism, this time including vector fields.  
In Section \ref{sformulas}
we  calculate the spectrum of $\zeta$ at tree and one-loop level, and the
bispectrum of $\zeta$ at tree level.  In Section \ref{sscalar} we recall the generation
of a scalar field perturbation from the vacuum. In Section \ref{stigauge}
we see how a gauge field perturbation can be generated.  In Section
\ref{smodgrav} we see how a vector field perturbation can be generated, using
a modified-gravity action without gauge invariance and including the 
longitudinal component. In Sections \ref{svc} and \ref{svi}
we see how a vector field perturbation can contribute to $\zeta$, through
respectively the vector curvaton and vector inflation mechanisms.
We conclude in Section \ref{scon}.

\section{Observational constraints on the curvature perturbation} 

\label{sobs}

Direct information on the curvature perturbation comes mostly from
measurements of 
the anisotropy of the CMB and the inhomogeneity of the galaxy distribution.  
These cover a limited range of scales, corresponding to roughly
$\Delta \ln k\sim 10$ where $k$ is the comoving wavenumber. 
Indirect information is available at much longer and shorter scales.
In this section we summarise the information.

\subsection{Formulas}

We are interested in the correlators of the curvature
perturbation, in particular the two-point correlator.
For any cosmological perturbation $\beta(\bfx)$, at some fixed time,
we define  Fourier components with normalisation
\be
\beta(\bfk) \equiv \int \beta(\bfx) e^{-i\bfk\cdot\bfx} d^3x
\,. \label{fouriercomp} \ee
Assuming that the  two-point correlator $\vev{\beta(\bfx)\beta(\bfx')}$
is invariant under translations (statistically  homogeneous),
the two-point correlator of the Fourier components takes the form
\be
\vev{\beta(\bfk)\beta(\bfk')}
= (2\pi)^3 \delta(\bfk+\bfk') \frac{2\pi^2}{k^3} \calp_\beta(\bfk) \,,
\label{spectrum} \ee
which defines the spectrum $\calp_\beta$ \footnote
{ The averages are  over 
some  ensemble of universes, of which our observable Universe is supposed to be
a typical realization.}.
 If the two-point correlator is also invariant under
rotations (statistical isotropy) the spectrum $\calp_\beta(\bfk)$
depends only on the magnitude $k$.
In that case we  shall sometimes invoke  a quantity
 $P_\beta(k) \equiv (2\pi^2/k^3) \calp_\beta(k)$.

By virtue of the reality condition $\beta(-\bfk)=\beta^*(\bfk)$, 
an equivalent definition of the spectrum is
\be
\vev{\beta(\bfk)\beta^*(\bfk')}
= (2\pi)^3 \delta(\bfk-\bfk') \frac{2\pi^2}{k^3} \calp_\beta(\bfk) \,.
\label{spectrum2} \ee
Setting $\bfk=\bfk'$ the left hand side is $\vev{|\beta(\bfk)|^2}$.
It follows that the  the spectrum is positive and nonzero.

Even if $\calp_\beta(\bfk)$  is anisotropic, the reality condition requires 
$\calp_\beta(\bfk)=\calp_\beta(-\bfk)$. The anisotropy will therefore be of the 
form \cite{acw} (see also Refs. \cite{ajian1,ajian2}) 
\be
\calp_\beta(\bfk) = \calp_\beta\su{iso}(k) \[ 1 + g_\beta
 (\hat{\bfd}\cdot \hat \bfk)^2 + \cdots \]
\,, \label{statan} 
\ee
where $\calp_\beta\su{iso}(k)$ is the average over all directions, $\hat{\bfd}$ 
is some
unit vector and $\hat \bfk$ is a unit vector along $\bfk$.

 If there is no
correlation between the Fourier components except for the reality condition,
the perturbation is said to be Gaussian. Then the two-point correlator
is given by \eq{spectrum} and 
the three-point correlator vanishes while the four-point correlator is
\begin{equation}
 \langle \beta_{\bfk_1} \beta_{\bfk_2} \beta_{\bfk_3} \beta_{\bfk_4} \rangle = 
\langle \beta_{\bfk_1} \beta_{\bfk_2} \rangle \langle \beta_{\bfk_3} \beta_{\bfk_4}
 \rangle +
\langle \beta_{\bfk_1}  \beta_{\bfk_3} \rangle \langle \beta_{\bfk_2} \beta_{\bfk_4}
\rangle + \langle \beta_{\bfk_1} \beta_{\bfk_4} \rangle \langle \beta_{\bfk_2}
\beta_{\bfk_3} \rangle \,.
\label{fourpoint} \end{equation}
The  five-point correlator vanishes and the six-point correlator
is given by the analogue of \eq{fourpoint}, and so on.
All correlators are known once the spectrum is specified. We conclude that
{\em a Gaussian perturbation is statistically homogeneous} even though it
need not be statistically isotropic.

Non-gaussianity is signalled by a non-vanishing 3-point correlator,
 an additional (`connected') contribution to the 4-point correlator and so on.
Statistical homogeneity requires that each correlator of Fourier components
vanishes unless the sum of the wave-vectors vanishes (generalising the
delta function of \eq{spectrum}), and statistical isotropy requires
that it is invariant under rotations. In particular, statistical homogeneity 
requires a  3-point correlator of  the form
\be
\vev{\beta(\bfk)\beta(\bfk')\beta(\bfk'')}
= (2\pi)^3 \delta(\bfk+\bfk'+\bfk'') B(\bfk,\bfk',\bfk'')
\,, \ee
and statistical isotropy requires that $B$ depends only on the magnitudes of
the vectors. Assuming statistical isotropy one also defines a reduced bispectrum
$\calb_\beta$ by
\be
B_\beta(k,k',k'') \equiv \calb_\beta(k,k',k'') \[ P_\beta(k) P_\beta(k') 
+ \mbox{\,cyclic permutations\,} \]
\,. \ee

\subsection{Spectrum and non-gaussianity}

Observational results concerning the spectrum $\calpz$ are generally
obtained with the assumption of statistical isotropy, but they would
not be greatly affected by the inclusion of anisotropy at the $10\%$ level.

Direct observation, coming from the anisotropy of the CMB and the 
inhomogeneity of the galaxy distribution, gives information on what are
called cosmological scales \cite{wmap5}. These correspond to a range
$\Delta \ln k\sim 10$ or so downwards from the scale $k\mone\sim H_0\mone$
that corresponds to the size of the observable Universe\footnote{As usual a subscript 0 indicates the present epoch, and we set $a_0=1$.}.
It is found that $\calpz$ is almost scale independent with the value
$\calpz\half \simeq 5\times 10^{-5}$. There is mild scale dependence
corresponding to
\be
n-1\equiv \frac{d\ln \calpz}{d\ln k} = -0.040 \pm  0.014
\,. \ee

On much bigger or smaller scales the constraint is far weaker.
Assuming a constant $n$ on such scales, they are
\be
-5 < n-1 \lsim  0.4 \frac {50} {N\sub{corr}}\,,\qquad N\sub{corr}
\equiv \ln(k\sub{corr}/k\sub{max})
\,. \label{extremen} \ee
The lower bound, referring to very large scales $k\ll H_0$, comes \cite{abook}
from the absence
of an enhancement of the CMB quadrupole (Grishchuk-Zeldovich effect).

The upper bound is more interesting. In this expression, $N\sub{corr}$
is the number of $e$-folds of inflation, between horizon exit for the
smallest cosmological scale $k\sub{max}\mone$ and horizon exit for the
smallest scale  $k\sub{corr}\mone$ on which the curvature perturbation exists
(correlation length). It corresponds \cite{klm}
to the following values for the spectrum
at those scales:
\be
\calpz\half(k\sub{max}) \lsim 5\times 10^{-5}\,,\qquad \calpz\half(k\sub{corr})
< 10\mone \,. \ee
The first number is the observed value on cosmological scales.
The second number corresponds to an order of magnitude  upper bound on 
the spectrum that under certain assumptions is required to avoid 
an overabundance of primordial black holes \cite{carrbh}.
Further discussion about the  upper bound on $\calpz$ is given in Ref. \cite{klm}.

If $\zeta$ is generated during inflation, or soon afterwards, $k\sub{corr}$
will be the scale leaving the horizon at the end of inflation.
Then $N\sub{corr}\simeq N-10$, where $N$ is the number of $e$-folds 
of inflation  after
the largest cosmological scale $H_0\mone$ leaves the horizon. 
For a high inflation scale and a fairly standard cosmology afterwards,
$N\simeq 60$ making $N\sub{corr}\simeq 50$. If instead $\zeta$ is formed
long after inflation, through say the curvaton model, $N\sub{corr}$ can be
much lower for the same $N$,
 and $N$ itself will be reduced if the inflation scale is low.

If the spectral 
tilt  varies, the upper bound  refers to average of the tilt  with respect
to $\ln k$, in the interval  $k\sub{max} <  k\sub{corr}$.
The possibility of large tilt  on small scales has been investigated in
Ref. \cite{klm}.
A strongly increasing tilt  on small scales  could come from a single mechanism
for generating $n$, such as the running mass inflation model. Alternatively,
a large and practically constant $n$ on small scales
 could be generated if  the curvature perturbation has two components:
\be
\calpz(k) = \calp\sub{flat}(k) + \calp\sub{steep}(k)
\label{flatsteep} \,. \ee
The first component might be nearly flat
and dominate on cosmological scales, while the second might have 
large tilt and dominate in the interval  $k\sub{max} < k <k\sub{corr}$.
In that case, the upper bound in Eq. \eqreff{extremen} applies to the 
spectral tilt of $\calp\sub{steep}$.

Coming to non-gaussianity, one generally focusses on the bispectrum, working 
with the quantity  $\fnl\equiv (5/6)\calb_\zeta$. 
If $\fnl$ is generated from  one or more gaussian field perturbations
with scale-independent spectra it is practically scale independent.
With that assumption, the most recent analysis \cite{ssz} finds
$\fnl=38\pm21$ at $1\sigma$ but $-4<\fnl< 80$ is allowed at $95\%$ confidence
level. For fully correlated
non-gaussianity,  $f\sub{NL}\calpz\mhalf$ is of order the fractional 
non-gaussianity of $\zeta$ which means that the non-gaussian fraction is
less than $10\mthree $ or so, and  in any case the observational bound on
$\fnl$ corresponds to a small non-gaussian fraction \cite{bl}.

Allowing scale dependence of the bispectrum, the observational bounds
are very weak on scales outside the cosmological range,
so that  for example
$\zeta$ could be the square of a gaussian quantity.

\subsection{Statistical anisotropy and statistical inhomogeneity}

Taking all the uncertainties into account, 
observation is consistent with statistical anisotropy and statistical
inhomogeneity but allows either of these things at around the $10\%$ level.
In this section we briefly review what is known.

Assuming statistical homogeneity of the curvature perturbation, 
a recent study \cite{ge} (see also Refs. \cite{pekowsky,samal}) of the cosmic microwave background radiation (CMB) 
temperature perturbation
finds weak evidence for
statistical anisotropy. They keep only the leading term of \eq{statan}:
\be
\calpz(\bfk) = \calp_\zeta\su{iso}(k) \( 1 + g
 (\hat{\bfd}\cdot \hat \bfk)^2 \) \,, 
\label{curvquad} \ee
and find $g\simeq 0.15\pm0.04$ with $\hat{\bfd}$ in a specified direction.
 The authors point out though that systematic uncertainties could  make $g$ 
compatible with zero. We will therefore just assume $|g|\lsim 0.3$ 
\footnote{A related work \cite{pullen} shows that the lowest detectable 
value for $|g|$ from the expected performance of WMAP is $|g| \simeq 0.1$. 
The same analysis gives the lowest detectable value from the expected 
performance of PLANCK: $|g| \simeq 0.02$.}.
In other words, we assume that the spectrum of the curvature perturbation
is isotropic to within thirty  percent or so.
There is at present no bound on statistical anisotropy of the 3-point
or higher correlators. 

In  some different studies, the mean-square CMB perturbation in opposite 
hemispheres has been measured, to see if there is any difference between
hemispheres.  A recent work \cite{dipole1,dipole2,dipole3} finds a 
difference of
order ten percent, for a certain choice of the hemispheres, with statistical
significance at the $99\%$ level. Given the difficulty of handling systematic 
uncertainties it would be premature to regard the evidence for this 
hemispherical anisotropy as completely overwhelming.

Let us see what hemispherical anisotropy would imply for the curvature
perturbation.
Focussing on a small patch of sky, the  statistical anisotropy of the 
curvature perturbation 
implies that the mean-square temperature perturbation within a 
given small patch will {\em in general} depend on the direction of that patch.
 This is because the mean square within such a patch depends
(in the sudden decoupling approximation) upon the mean square of the 
curvature perturbation in a small planar region of space
 perpendicular to the line of sight located at 
last scattering\footnote{The  sudden decoupling is not essential here. 
It can be replaced by the
exact line of sight formalism, leading to the same conclusion.}.
But the mean-square temperature
 will be {\em the same} in patches at opposite
directions in the sky, because they explore the curvature perturbation
$\zeta(\bfk)$ in the same $\bfk$-plane and the spectrum $\calpz(\bfk)$
is  invariant under the change $\bfk\to-\bfk$. It follows that statistical
anisotropy of the curvature perturbation cannot by itself generate a 
hemispherical anisotropy.

In the above  discussion of the CMB temperature perturbation, we ignored cosmic
variance, by identifying the measured mean-square temperature perturbation
within a given patch with the ensemble average of that quantity.
That will certainly be permissible if the multipoles of the CMB, 
{\em including the lowest ones}, are almost
uncorrelated corresponding to an almost gaussian
curvature perturbation. 

With the caveat concerning cosmic variance,  we conclude that hemispherical 
anisotropy of
the CMB temperature requires statistical {\em inhomogeneity} of the 
curvature perturbation. Then 
$\vev{\zeta(\bfk)\zeta({\bfk'})}$ is not proportional to $\delta(\bfk+\bfk')$.
But in a small region of the observable Universe it might still be reasonable
to invoke approximate statistical homogeneity, by 
 defining  a 
position-dependent spectrum $\calp(k,\bfx)$ (taken for simplicity to be 
rotationally invariant). This way of generating the hemispherical anisotropy
has been considered in Refs. \cite{chrisg,erickcek}, 
but is outside the framework of the present paper. 

Before ending this section we note that, in 
addition to the primordial curvature perturbation, there might be a primordial
tensor perturbation with  spectrum  $\calp_h$ \cite{abook}. The fraction 
$r\equiv\calp_h/\calpz$ is constrained by observation to be $\lsim 0.1$ \cite{wmap5}.

\section{The $\delta N$ formalism} 

\label{sdeln}

The $\delta N$ formalism for scalar field perturbations was
given at the linear level in Refs. \cite{starobinsky,ss}. At the non-linear
level which generates  non-gaussianity  
it was described in Refs. \cite{lms,DY}. 
Here we extend the formalism to include vector fields.

With generic coordinates the line element of the perturbed universe is
\be
ds^2 = g_\mn dx^\mu dx^\nu \,.
\ee
The coordinate system of the perturbed universe defines a slicing 
(constant time coordinates) and a threading (constant space
coordinates) of spacetime. 

To define the cosmological perturbations, one chooses a coordinate
system in the perturbed universe,  and then compares that universe  with an 
unperturbed one. The unperturbed universe is taken to be
homogeneous, and  is usually taken to be isotropic as well. 
  In this Section though, we develop the $\delta N$
formalism without assuming isotropy.

The $\delta N$ formalism does not invoke a theory of gravity, but it
does invoke an energy-momentum tensor $T_\mn$. {}From a mathematical viewpoint,
any definition will do provided that it  satisfies the 
continuity equation $\bigtriangledown_\mu T^\mu_\nu=0$ with $\bigtriangledown_\mu$ the covariant derivative.
Following for instance Refs.
\cite{lwconserved,VI},   we define  $T_\mn$ in terms of the 
spacetime curvature:
\be
 R_\mn  - \frac12 g_\mn R  =  - 8\pi G T_\mn \,. \label{tmndef1}
\ee
This is the Einstein field equation if,
in a locally inertial frame,  $T_\mn$ is the energy-momentum tensor of
Special Relativity. In the context of field theory, this means that the
action should be of the form
\be
S = \int d^4x \sqrt{-g} \[ \frac12 m_P^2 R + \call \] \,, \label{generals}
\ee
where $m_P \equiv (8\pi G)^{-1/2}$ is the reduced Planck mass, and  $\call$, evaluated in a locally inertial frame, 
is the lagrangian density
of flat spacetime field theory.  Then
$T_\mn$ is the `improved energy-momentum tensor' which is given
in terms of the fields by a standard expression. The bosonic part
$\call\sub{bos}$ of $\call$ gives a contribution
\be
T_\mn\su{bos}  = 2 \frac{\pa \call\sub{bos}}{\pa g^\mn} 
- g_\mn \call\sub{bos} \,. \label{tmndef}
\ee

Of course, we can always write the action in the form given by Eq. \eqreff{generals}
with some $\call$. When that is done, the contribution of the bosonic part
$\call\sub{bos}$ will still be given by \eq{tmndef}. 
We shall invoke this expression in several cases where Einstein gravity holds,
and will invoke it in Section \ref{svi} for a case where Einstein gravity
does not hold, dropping the label `bos'.

\subsection{The curvature perturbation and the tensor perturbation}

To define the curvature perturbation, we 
smooth the metric tensor and the energy-momentum tensor on 
a comoving scale $k\mone$
significantly shorter than the scales of interest,
and we consider the super-horizon regime $aH\gg k$.
 On the reasonable assumption that the smoothing scale is the biggest relevant
scale,  spatial gradients of the smoothed 
metric and energy-momentum tensors will
be negligible. 
As a result,  the evolution  of these quantities at
each comoving location will be that of some homogeneous `separate universe'. 
In contrast with earlier works on the separation universe assumption,
we will in this section allow the possibility that the separate
%paper
universes are anisotropic even though homogoneous.

We consider the  slicing
of spacetime with uniform energy density, and the  threading which moves
with the expansion (comoving threading). 
By virtue of the separate universe assumption, the 
threading will be orthogonal to the slicing.
 The spatial metric  can then be written as
\be
g_{ij}(\bfx,\tau) \equiv a^2(\bfx,\tau) \(Ie^{2h(\bfx,\tau)} \)_{ij}
\,, \label{gij} \ee
where  $I$ is the unit matrix, and the matrix
$h$ is traceless, which means that 
 $Ie^{2h}$  
has unit determinant.  The time dependence of the locally defined scale factor 
$a(\bfx,t)$ defines the rate at which an infinitesimal
comoving  volume $\calv$ expands:
$\dot \calv/\calv=3\dot a/a$.  

We split $\ln a$ and $h_{ij}$ into an unperturbed part plus a perturbation:
\bea
\ln a(\bfx,\tau) &\equiv& \ln a(\tau) + \zeta(\bfx,\tau)\,, \\
h_{ij}(\bfx,\tau) &\equiv& h_{ij}(\tau) + \delta h_{ij}(\bfx,\tau)
\,. \eea
The unperturbed parts can be defined as spatial averages within the 
observable Universe,
but any definition will do as long as it makes the perturbations small 
within the observable Universe. If they are small enough,
 $\zeta$ and $\delta h_{ij}$ can be treated as first-order perturbations. 
That is expected
to be the case, with the proviso that a second-order treatment of 
$\zeta$ will be necessary
to handle its non-gaussianity if that is present at a level corresponding 
to $\fnl\lsim 1$
(with the gaussian and non-gaussian components correlated) \cite{lr1}.

\subsubsection{The curvature perturbation}

In this paper we are mainly concerned with 
 the curvature perturbation $\zeta$ \footnote{It is so-called because one usually
has in mind the case that $\delta h_{ij}$ is negligible; of course it too
corresponds to a perturbation in the spatial curvature.}. 
%We will employ the $\delta N$ formula, which
%makes no reference to cosmological perturbation theory and is valid no matter how
%small is the non-gaussianity of $\zeta$.
Because $I e^h$ has unit determinant, the 
 energy continuity
equation $d(\calv \rho)= - P d\calv$ implies that $\dot\zeta$ is independent
of position, 
during any  era when the pressure $P$ 
 is a unique function of the energy density $\rho$ \cite{lms} (hence uniform 
on slices
of uniform $\rho$).
%\footnote
%{As far as the mathematics goes, any
% definition of $\rho$ and $P$ will do a long as they satisfy the
%energy continuity equation. If the theory of gravity satisfies the 
%equivalence principle (in particular, if Einstein's theory is valid)
% one should use the flat spacetime definitions
%in a locally inertial frame. Otherwise one might 
%take the stress-energy tensor to be defined by the Einstein field equation,
%even though that equation will not have its usual significance. 
%The latter definition
%of the stress-energy tensor is used in  Ref. \cite{VC}.}.
  Absorbing $\dot\zeta$ into the unperturbed scale
factor,  $\zeta(\bfx)$ is then time independent. 

From the success of
Big Bang Nucleosythesis, we know that Einstein gravity is a
good approximation
%when the Universe
when the shortest  cosmological scale approaches
 horizon entry at $T\sim 1\MeV$. Also, the cosmic fluid is then
 radiation dominated to high accuracy 
implying $P=\rho/3$ and a constant value of $\zeta$. 
We denote this value simply by $\zeta(\bfx)$, and it is the one constrained by
observation as described in Section \ref{sobs}.

\subsubsection{The tensor perturbation}

\label{stp}

The perturbation $\delta h_{ij}$ may also be of interest. We discuss it at this point
in general terms, and in Section \ref{svi} we  provide an  explicit calculation  
within the vector inflation model. 

Consider first the unperturbed quantity $h_{ij}(\tau)$.   
In this paper we  are taking the unperturbed expansion to be 
practically isotropic expansion with Cartesian coordinates. As a result, 
 we can take the unperturbed quantity 
 to vanish so that  $a(\tau)$ is the unperturbed scale factor. 
More generally, if  the unperturbed quantity is  any time-independent 
 matrix, we can make a linear coordinate transformation
which diagonalises $Ie^h$ and can then choose the normalization of the
scale factor so that $h_{ij}$ again vanishes. A time-dependent
unperturbed quantiy $h_{ij}(\tau)$ would correspond to an 
 unperturbed Universe with   anisotropic expansion. 

If one or more   vector fields exist during inflation, one might think that 
the unperturbed expansion may easily be anisotropic. Assuming Einstein gravity 
though, that is not the case because according to a theorem of Wald  \cite{wald,wald2}
enough inflation driven by a constant scalar field potential 
will isotropise the expansion\footnote{He calls this constant potential a 
cosmological constant.}.
This statement becomes only an approximation for realistic slow roll inflation where
the potential is varying, and it doesn't apply to `vector inflation' models
where  inflation is driven by a
constant vector field potential  \cite{ford,triplet,VI,armendariz}. 
For vector inflation
though, one can ensure approximate isotropy of the expansion by invoking a large
number of independent fields \cite{VI}, as we shall discuss in Section
\ref{svi}. 
%Discounting  vector inflation and Assuming   Einstein gravity, 
%We  therefore expect  the back reaction of
%unperturbed vector fields, on the metric during slow roll scalar field inflation,
%to be  very small though perhaps not entirely negligible \cite{kksy,watanabe}.

After inflation, an era of anisotropic stress (from vector fields or any other source)
might  cause significant
anisotropy of the expansion, but that does not happen in the usual scenarios of
the early Universe. Assuming Einstein gravity, the anisotropy
will anyway decay when the anisotropic stress switches off. 

As we are dealing with a smoothed metric well after
horizon exit, the status of the perturbed quantity $h_{ij}(\bfx,\tau)$ 
at  a given location is the same as that of the unperturbed quantity. 
The anisotropy of the local expansion will be negligible if Wald's theorem holds
or if there is vector inflation with a sufficiently large number of independent
vector fields. Then 
the  perturbation $\delta h_{ij}(\bfx,\tau)$ is almost
time independent.  Also, we expect $\delta h_{ij}$ to remain
time-independent after inflation 
until the approach of horizon entry, since the local anisotropic stress
is expected to remain negligible.

Now we consider first order cosmological perturbation theory, taking the unperturbed
$h_{ij}$ to vanish. At first order,
the equations satisfied by the cosmological 
perturbations comprise three uncoupled modes, termed
scalar, vector and tensor.
The first order perturbation
$\delta g_{ij}$ is equal to $\delta_{ij}\zeta+ \delta h_{ij}$ with $\zeta$
belonging to the scalar mode. Setting spatial gradients equal to zero in accordance
with the separate universe assumption, $\delta h_{ij}$ satisfies the transversality
condition $\pa_i \delta h_{ij}=0$, which means that it  belongs to the tensor mode.
It can be written in terms of polarization tensors as $e^+_{ij}h_++e^-_{ij}h_-$,
and assuming statistical parity invariance each of the amplitudes has the same
spectrum $\calp\sub{ten}/4$. 
%Assuming Einstein gravity and negligible anisotropic stress, its constant value
%$\delta h_{ij}$ is constrained by observation.  Its spectrum as a 
The fraction $r\equiv \calp\sub{ten}/\calpz$ 
is $\lsim 10\mone$ \cite{wmap5} and future 
measurements will reduce this bound
by a factor of 10 to 100, or detect $r$ \cite{baumann}.
 
Let us discuss the origin of  the tensor perturbation 
$\delta h_{ij}$, within first order cosmological perturbation
theory assuming Einstein gravity. 
The standard calculation assumes negligible anisotropy in the inflationary expansion
and  negligible anisotropic stress (which will certainly be the case if
scalar fields dominate). Under these assumptions, each of $h_{+,\times}/\sqrt2 m_P$ 
has the action of a free scalar field. The  classical equation of motion is
\be
\ddot \delta h_{ij} + 3H \dot \delta h_{ij} + (k^2/a^2) \delta h_{ij} = 0
, \label{hijeqn} \ee
which makes $\delta h_{ij}$ constant after horizon exit. The 
 spectrum of the perturbation generated from the vacuum fluctuation is
\be
\calp\sub{ten} =\frac 8{\mpl^2} \( \frac{H_*}{2\pi} \)^2
\label{calpten}
, \ee
%  is generated from a vacuum fluctuation, and taking the inflationary energy density
%to have a constant value $\rho_*$ its spectrum is given by  $r=(\rho_*\quarter/
%3.3\times 10^{16}\GeV)$ 
which is too small to observe in small-field inflation models.
%models. The standard calculation assumes isotropic expansion though,
%corresponding to a time independent $\delta h_{ij}$. 

If one or more vector fields is relevant during inflation, the unperturbed 
expansion will be anisotropic at some level. Then, even if the vector field is 
unperturbed,  $\delta h_{ij}$  will time dependent during inflation, 
%and generate an observable $\delta h_{ij}$ 
and can be generated even if the  vacuum fluctuation is negligible; such a contribution
would be  correlated with the curvature perturbation \cite{kksy,watanabe}, in contrast
with the one generated from the vacuum fluctuation. A perturbation of the vector
field will give an additional effect. As we argued earlier these effects are expected
to be neligible if Einstein gravity holds, assuming that 
 either Wald's theorem applies or there is vector  inflation with sufficiently many
independent fields.

 Within this first order treatment, the tensor perturbation is
gaussian. Since the tensor perturbation has yet to be detected there is little 
motivation
to consider its non-gaussianity. At the time of writing, the only calculation  of 
non-gaussianity  has been done by Maldacena \cite{maldacena}
assuming single field slow roll inflation with Einstein gravity. Using second  order
perturbation theory he chooses a gauge where $\delta h_{ij}$ is transverse as well
as traceless. He calculates the three-point correlators  involving Fourier
components of $\zeta$ and/or $\delta h_{ij}$, at the epoch soon after
horizon exit,  and finds them to be suppressed by
slow roll factors. If $\zeta$ receives contributions only from the inflaton
perturbation, it is constant after horizon exit and then the three point correlator
of $\zeta$ corresponds to  $\fnl\sim 10\mtwo$ which is almost certainly 
too small ever to detect. There is no reason to think that the correlators involving
$\delta h_{ij}$ will be detectable either. 
Judging by this example, there is no need for
the discussion of $\delta h_{ij}$ to 
go beyond first order cosmological perturbation theory.

\subsection{The $\delta N$ formula}

Keeping the comoving threading, we can write the analogue of 
 \eq{gij} for a generic slicing:
\be
\tilde g_{ij}(\bfx,\tau) \equiv \tilde a^2(\bfx,\tau) \(Ie^{2\tilde h(\bfx,\tau)} \)_{ij}
, \label{gijtilde} \ee
with again  $Ie^{2\tilde h}$  having unit determinant so that 
the rate of volume expansion is $\dot \calv/\calv = 3\tilde a(\bfx,t)$.
Starting with an initial `flat' slicing such that the locally-defined scale
factor is homogeneous, and ending with a slicing of uniform density, 
 we then have 
\be
\zeta(\bfx,t) = \delta N(\bfx,t)
, \ee
where the number of $e$-folds of expansion is defined in terms of the volume
expansion by the usual expression $\dot N = \dot \calv/3\calv$.
The choice of the initial epoch has no effect on $\delta N$, because
the expansion going from one flat slice to another is uniform.
We will choose the  initial epoch to be 
 a few Hubble times after the smoothing
scale leaves the horizon during inflation.
According to the usual assumption, the 
evolution
of the local expansion rate  is determined by the initial values
of one or more of the perturbed  scalar fields $\phi_I$. 
Then we can write
\bea
\phi_I(\bfx) &=& \phi_I + \delta\phi_I(\bfx), \\
\zeta(\bfx,t) &=& \delta N(\phi_1(\bfx),\phi_2(\bfx),\ldots,t) \nonumber \\
&=&  N_I(t) \delta\phi_I(\bfx) + \frac12 N_{IJ}(t)
\delta\phi_I(\bfx) \delta\phi_J (\bfx) + \ldots , \label{dNsc}
\eea
where $N_I\equiv \pa N/\pa \phi_I$, etc.,\ and the partial derivatives
are evaluated with the fields at their unperturbed values denoted
simply by $\phi_I$. The field perturbations $\delta \phi_I$ in Eq. (\ref{dNsc}) are defined on the `flat' slicing
such that $a(\bfx,t)$ is uniform. 

The unperturbed field values 
are defined as the spatial averages, over a  comoving
box within which the perturbations are defined. The box size $aL$
should satisfy $LH_0\gg 1$ so that the observable
Universe should fit comfortably inside it \cite{mybox}. 
If there have been exponentially many $e$-folds of inflation before  the observable
Universe leaves the horizon, one could choose
 $\ln(LH_0)$ to be exponentially large, but that would not be a good idea because
it introduces unknowable new physics and places the calculation out of control
\cite{mybox}. One therefore chooses a `minimal box', such that $\ln(LH_0)$
is significantly bigger than 1 without being exponentially large.

The spatial averages of the scalar fields, that determine $N_I$, etc., and hence
$\zeta$ cannot in general be calculated. Instead they are parameters, that have
to be specified along with the relevant parameters of the action before the 
correlators of $\zeta$ can be calculated. The only exception is when $\zeta$
is determined by the perturbation of the inflaton in single-field
inflation. Then, the unperturbed field value when cosmological scales leave
the horizon can be calculated, knowing the number of $e$-folds to the end of
inflation which is determined by the evolution of the scale factor after 
inflation.
Although the unperturbed field values cannot be calculated, their mean square
for a random location of the minimal box (ie.\ of the observable Universe)
can sometimes be calculated using
the stochastic formalism \cite{stochastic}.

In this paper we suppose that one
or more  perturbed vector fields also  affect 
the evolution of the local expansion rate. 
Keeping for  simplicity one scalar field and one vector field we have
\be
\zeta(\bfx,t) = \delta N(\phi(\bfx),A_i(\bfx),t) =  N_\phi \delta\phi + N_A^i \delta A_i
+ \frac12 N_{\phi \phi}(\delta\phi)^2 + 
\frac12 N_{\phi A}^i\delta\phi \ \delta A_i +  
\frac12 N_{AA}^{ij}\delta A_i \ \delta A_j + ... \,, \label{series}
\ee
where 
\be
N_\phi\equiv\frac{\partial N}{\partial \phi}\,,\quad
N_A^{i}\equiv\frac{\partial N}{\partial A_i}\,,\quad
N_{\phi\phi} \equiv\frac{\partial^2 N}{\partial \phi^2}\,,\quad
N_{AA}^{ij}\equiv\frac{\partial^2 N}{\partial A_i\partial A_j}\,,
\quad N_{\phi A}^i\equiv\frac{\partial^2 N}{\partial A_i\partial\phi}\,,
\ee
with $i$ denoting the spatial indices running from 1 to 3.
As with the scalar fields, the 
unperturbed vector field values are defined as averages within the chosen
box.

In these formulas  there is no need to define the basis (triad) for
the components $A_i$. Also, we need not assume that $A_i$ comes from a 4-vector
field, still less from a gauge field.

The discussion so far allows the unperturbed expansion to be anisotropic.
In the following sections though, we will take it to be isotropic.
Also, we take the unperturbed spatial geometry to be flat. Then the
unperturbed   line element is
\be
ds^2 =a^2(\tau) \(  -d\tau^2 +  \delta_{ij} dx^i dx^j \)
, \label{rw} \ee
where $\tau$ is conformal time and $a$ is the scale factor. 
Depending on the context, we may instead  use cosmic time $t$ corresponding to
$dt=  a d\tau$. 
%The vector field  lives in flat space. 
We shall take  $A_i$ to be the physical field,
defined with respect to the  orthonormal basis induced by the
Cartesian space coordinates $r^i=a(t) x^i$. 
We shall also have occasion to consider the field $B_i=a A_i$ 
that is defined with
respect to the orthogonal (but not orthonormal) basis induced by the
comoving coordinates $x^i$. The corresponding upper-index
quantities are $A^i=A_i$ and $B^i = a\mtwo B_i$. 

\subsection{The growth of $\zeta$}

As noted earlier, $\zeta$ is constant during any era when pressure $P$ is a 
unique function of energy density $\rho$. 
In the simplest scenario, the field
whose perturbation generates  $\zeta$ is the inflaton field $\phi$ in a single-field
model. Then the local value of $\phi$ is supposed to determine the subsequent
evolution of both pressure  and  energy density, 
 making $\zeta$ constant from the beginning.

Alternatives to the simplest 
 scenario generate all or part of $\zeta$ at successively
later eras. Such generation is possible during any era,  unless there is sufficiently
complete matter domination ($P=0$) or radiation domination ($\rho=P/3$).
Possibilities in chronological order include generation during 
(i) multi-field inflation  \cite{starobinsky},  (ii)
 at the end of inflation \cite{myend}, (iii) during preheating, 
(iv) at reheating, and (v) at a
second reheating through the curvaton mechanism 
\cite{curvaton1,curvaton2,curvaton3,luw}.

A vector field cannot replace the scalar field in the simplest scenario,
because unperturbed 
inflation with a  single unperturbed vector field will be  very 
anisotropic and so will be the resulting curvature perturbation. Even with isotropic
inflation, we are about to see that a  single 
vector field perturbation cannot  be responsible for the entire curvature
perturbation (at least in the scenarios that we discuss)
 because its contribution is 
%likely to be 
highly anisotropic.
It could instead be responsible for part of the curvature perturbation,
through any of the mechanisms listed above. Of these, the end of inflation
mechanism 
has already been explored \cite{ys}. In this paper we explore another
one, namely the vector curvaton mechanism \cite{VC}. We will also explore the 
vector inflation scenario \cite{VI}, according to which inflation
is driven by a large number of randomly oriented vector fields which can give
sufficiently isotropic inflation and (as we shall see) an extremely 
isotropic $\zeta$.

\section{Formulas for the spectrum and bispectrum of the curvature perturbation}

\label{sformulas}

\subsection{Spectrum of the vector field perturbation}

In Section \ref{sscalar} we describe the standard scenario for generating the
 scalar field perturbations from the vacuum. Within this scenario,
these perturbations are  Gaussian with no correlation between different
perturbations. Their  stochastic properties are defined by the
spectrum $\calp_{\delta\phi}$ of each field.
Either of the equivalent definitions  \eqsref{spectrum}{spectrum2} can be
used to define the spectrum, with 
 $\beta=\delta\phi$.

To  deal with  a  vector field perturbation $\delta A_i$ we write   
\be
\delta A_i (\bfk,\tau) \equiv  \sum_\lambda e^\lambda_i(\hat{\bfk}) \delta
A_\lambda(\bfk,\tau)
\,,
\ee
where with the $z$ axis along $\bfk$ the polarization vectors are defined by
\be
e^L \equiv (1,i,0)/\sqrt{2}\,,\qquad e^R \equiv (1,-i,0)/\sqrt{2}\,,\qquad
e^{\rm long}  \equiv (0,0,1)
\,. \label{polvecs} \ee
 These expressions define the polarization vectors only up to a rotation 
about the $\bfk$ direction but that is enough for the present purpose.
We will let   the change $\bfk\to-\bfk$
reverse $z$ and $x$ but not $y$. Then $e_\lambda(-\hat \bfk)
=-e_\lambda^*(\hat \bfk)$
and there is a reality condition $A_\lambda^*(\bfk,\tau)=
-A_\lambda(-\bfk,\tau)$.

If the vector field corresponds to a gauge field, we  choose the gauge
so that $A\sub{long}=0$ leaving only $A_L$ and $A_R$. Otherwise we have to
keep all three $A_\lambda$.

In Sections  \ref{stigauge} and \ref{smodgrav} we describe two  scenarios for
generating the vector field perturbations $\delta A_\lambda$.
 Within both of them,  these  perturbations  are statistically isotropic
and Gaussian, with 
no correlation between different $\lambda$ or between the perturbations
of different fields (scalar or vector). As a result we need only to 
consider   the spectra $\calp_\lambda \equiv \calp_{\delta A_\lambda}$.
They can be defined by the analogue of either \eq{spectrum} or \eq{spectrum2}:
\bea
\vev{\delta A_\lambda(\bfk)\delta A_\lambda^*(\bfk')}
&=& (2\pi)^3 \delta(\bfk-\bfk') \frac{2\pi^2}{k^3} \calp_\lambda(k) \,, \\
\vev{\delta A_\lambda(\bfk)\delta A_\lambda(\bfk')}
&=& -(2\pi)^3 \delta(\bfk+\bfk') \frac{2\pi^2}{k^3} \calp_\lambda(k) 
\,. \eea
The spectra are nonzero and positive, with the minus sign in the second 
expression coming from $e_\lambda(-\hat \bfk)=-e_\lambda^*(\hat \bfk)$.

We will normally have  $\calp_L=\calp_R$, since 
a difference between these quantities would indicate parity violation of the
evolution of $A_i$. It is therefore useful to define
\be
\calp_\pm \equiv \frac12 \( \calp_R \pm \calp_L \) \,,
\ee
so that only $\calp_+$ will normally be present\footnote{Calculations that generate 
$\calp_-$ as well are  described in Refs. \cite{bamba,campanelli}.}.

% As we shall see
In the models that we discuss,  
the scale dependence of the spectra $\calp_\lambda(k)$ 
comes from the evolution of the perturbation $\delta A_\lambda$ after horizon
exit during inflation. In this regime, the spatial gradient
$k/a$ is  negligible compared with the Hubble parameter, and we expect that
it will be negligible compared with any other relevant parameter\footnote{This is 
verified for the specific scenarios that we consider.}.
In that case, the evolution of  $\delta A_\lambda(\bfx,\tau)$ at each position
will be the same as for the unperturbed field $A_i(\tau)$. By rotational
invariance the evolution of the latter is independent of $i$. Therefore,
we expect that the evolution of the three perturbations $\delta A_\lambda$
will become the same after horizon exit, giving them the same spectral index.
In that case $r\sub{long}$, defined as $r\sub{long} \equiv \calp\sub{long}/\calp_+$, 
will be just a number,  independent of $k$.

The correlators  of the $\delta A_i(\bfk)$ are
\be
\vev{\delta A_i(\bfk) \,\delta A_j(\bfk') } = (2\pi)^3 \delta(\bfk+\bfk') 
\frac{2\pi^2}{k^3} \Big[ 
T_{ij}^{\rm even} (\bfk) \calp_+(k)  +
i T_{ij}^{\rm odd} (\bfk) \calp_-(k)  + T\su{long}_{ij} (\bfk) \calp_{\rm long}(k) 
\Big] \,,
\ee
where
\be
T_{ij}\su{even} (\bfk) \equiv \delta_{ij} -  \hat k_i \hat k_j,\qquad
T\su{odd}_{ij} (\bfk) \equiv \epsilon_{ijk}\hat k_k,\qquad
T\su{long}_{ij} (\bfk) \equiv \hat k_i \hat k_j \,. 
\ee

\subsection{Spectrum of $\zeta$} \label{pz}

\subsubsection{Tree-level spectrum}

\label{streespec}

Since $\zeta$ is gaussian to high accuracy, 
 it seems reasonable to  expect 
that $\zeta$ will be  dominated
by one or more of the linear terms in \eq{series}. Keeping only them (corresponding to 
what is called the tree-level contribution) we  find\footnote
{The terminology tree-level and  one-loop corresponds to a Feynman graph
formalism \cite{bksw} that could easily be extended to include vector
fields.} 
\bea
\calp_\zeta^{\rm tree}(\bfk) &=& N_\phi^2 \calp_{\delta \phi}(k) + 
N_A^iN_A^j\Big [ 
T\su{even}_{ij} (\bfk) \calp_+(k) +
 T\su{long}_{ij} (\bfk) \calp_{\rm long}(k) \Big]  \label{Pzeta}\\
&=&
 N_\phi^2 \calp_{\delta \phi}(k) + 
N_A^2 \calp_+(k) + (\bfn_A\cdot \hat\bfk )^2 \calp_+(k)  \( r\sub{long} - 1  \)
  \label{Pzeta2}
\,. \eea
%where $r\sub{long} \equiv \calp\sub{long}/\calp_+$.
The above corresponds to \eq{curvquad} 
with $\hat \bfd=\hat \bfn_A$, $\bfn_A$ being the Cartesian vector with components $N_A^i$, and 
\bea
\calpz\su{iso}(k) &=& N_\phi^2 \calp_{\delta \phi}(k) + N_A^2
 \calp_+(k)\,, \label{PfA} \\
g &=&  \( r\sub{long}-1 \) \frac{ N_A^2 \calp_+(k) }
{ N_\phi^2 \calp_{\delta \phi}(k) + N_A^2 \calp_+(k) }
\,, \eea
where $N_A \equiv \sqrt{N_A^i N_A^i}$ is the magnitude of $\bfn_A$.
The   spectrum  is  scale-invariant if the spectra of the field
perturbations are  scale invariant. 

If the vector field perturbation  dominates $\zeta$ we have simply
$g = r\sub{long} - 1$. If the vector field is a gauge field $r\sub{long}=0$,
and if its  action is \eq{L} below $r\sub{long}=2$.
In both cases, the  the observational  bound $|g|\lsim 0.3$ is violated
which means that the vector field contribution cannot dominate.
If  there is no other vector
field contribution, the dominant contribution to $\zeta$ must then come
from one or more scalar field perturbations.

To avoid the need for scalar perturbations, one can suppose that  a
 large number $\caln$ of   vector fields perturbations 
 contribute to $\zeta$, with random orientation of the unperturbed fields.
With a sufficient number of fields, there is then no
 preferred direction and the  curvature
perturbation is isotropic.  

\subsubsection{One-loop contribution}

Using \eq{fourpoint}, 
the  contribution from the quadratic terms (one-loop contribution) is
\bea
\calp_\zeta^{\rm 1-loop} (\bfk) &=& \int 
\frac{dp \ p^2 k^3}{|\bfk + \bfp|^3 p^3} 
\Big \{ \frac{1}{2} N_{\phi \phi}^2  \calp_{\delta \phi} (|\bfk + \bfp|) 
\calp_{\delta \phi}(p) + \nonumber \\
&& + \frac{1}{4} N_{\phi A}^i N_{\phi A}^j \calp_{\delta \phi} (|\bfk + \bfp|) 
\Big[
T\su{even}_{ij} (\bfp) \calp_+(p) + T\su{long}_{ij} (\bfp) \calp_{\rm long} (p) 
\Big] + \nonumber \\
&& + \frac{1}{2} N_{AA}^{ij} N_{AA}^{kl} \Big \{
T\su{even}_{ik} (\bfk + \bfp) T\su{even}_{jl} 
(\bfp)  \calp_+ (|\bfk + \bfp|)  \calp_+ (p) + \nonumber \\
&& + 
T\su{odd}_{ik} (\bfk + \bfp) T\su{odd}_{jl} 
(\bfp)  \calp_- (|\bfk + \bfp|)  \calp_- (p) + \nonumber \\
&& + T\su{long}_{ik} (\bfk + \bfp) T\su{long}_{jl} (\bfp)
\calp_{\rm long} (|\bfk + \bfp|)  \calp_{\rm long} (p) + 
\nonumber \\
&& + 2 T\su{even}_{ik} (\bfk + \bfp) T\su{long}_{jl} (\bfp) 
\calp_+ (|\bfk + \bfp|)  \calp_{\rm long} (p) \Big \} 
\Big \} \,. \label{Pzetal}
\eea

If the spectra are scale-independent, the integral is proportional to 
$\ln(kL)$ \cite{myaxion} where  $L$ is the box size. If we allow
$\ln(kL)$ to be exponentially large the one-loop contribution can dominate
the tree-level contribution even with $\zeta$ almost gaussian, 
 but the whole calculation is then out of 
control \cite{mybox}. With a `minimal' box size such that $\ln(kL)$ is
not exponentially large, and keeping only a single scalar field contribution,
it has been shown \cite{mybox}
that the  ratio $(\calpz\su{1-loop}/\calpz\su{tree})\half$
is  of order the fractional non-gaussianity 
$f\sub{NL}\calpz\half$ of the curvature
perturbation which from observation is $\lsim 10\mthree$. 
% Since $\zeta$ is known to be gaussian to high accuracy, it must be
%dominated by one or more tree-level terms \cite{mybox}.
However, the loop contribution to $\zeta$ 
from a {\em given}  field could dominate the
tree level from that field, if both contributions
are  small compared with the total.
This could in particular be the case for the vector field contribution.

\subsection{Bispectrum of $\zeta$} \label{bz}

Working to leading order in the quadratic terms of the $\delta N$ formula,
we arrive at the   tree-level contribution to the bispectrum. Evaluating
it using \eq{fourpoint} we find
\bea
B_\zeta^{\rm tree} (\bfk,\bfk',\bfk'') &=& N_\phi^2 N_{\phi \phi} 
[P_{\delta \phi} (k) P_{\delta \phi} (k') + {\rm cyc. \ perm.}] + \nonumber \\
&& +  \frac{1}{2} N_\phi N_A^i N_{\phi A}^j \Big\{P_{\delta \phi} (k) \Big[
T_{ij}^{\rm even} (\bfk') P_+ (k') + 
i T_{ij}^{\rm odd} (\bfk') 
P_- (k') + T_{ij}^{\rm long} (\bfk') P_{\rm long} (k')\Big] + 5 \ {\rm perm.} 
\Big\} + \nonumber \\
&& + N_A^i N_A^j N_{AA}^{kl} \Big\{\Big[
T_{ik}^{\rm even} (\bfk) P_+ (k) + 
i T_{ik}^{\rm odd} (\bfk) P_- (k) + T_{ik}^{\rm long} (\bfk) P_{\rm long} (k)\Big]
 \times \nonumber \\
&& \times \Big[
T_{jl}^{\rm even} (\bfk') P_+ (k') + 
i T_{jl}^{\rm odd} (\bfk') P_- (k') + T_{jl}^{\rm long} (\bfk') P_{\rm long} 
(k')\Big] + {\rm cyc. \ perm.} \Big\} \,, \label{Bzeta}
\eea
where 
$P_{\delta \phi} (k)$ and $P_\lambda (k)$ are defined as
\be
P_{\delta \phi} (k) = \frac{2\pi^2}{k^3} \calp_{\delta \phi} (k) \,,\qquad
P_\lambda (k) = \frac{2\pi^2}{k^3} \calp_\lambda (k) \,. 
\ee

Reversal of the three wave-vectors corresponds to the parity transformation,
and from the reality condition $\zeta(-\bfk)=\zeta^*(\bfk)$
it changes each correlator into its complex
conjugate. For the spectrum  this is not of interest because
the reality condition also makes the spectrum real. For the 
bispectrum {\em with statistical isotropy}  it is also not of interest,
because the reality condition plus statistical isotropy 
make the bispectrum real\footnote{The triangle of vectors obtained by reversing
the vectors can be brought into coincidence with the original triangle by
a rotation.}. In our case, the bispectrum is statistically  anisotropic,
and  is guaranteed to be real only if the parity-violating spectrum 
$\calp_-$ vanishes.

Existing analysis of the bispectrum assumes statistical
isotropy\footnote{See for instance
Ref. \cite{babich}.}, and it seems important that the  analysis  should be
extended to allow for anisotropy and possible parity violation. 
The relation between non-gaussianity and the anisotropy of the spectrum
is explored in Ref. \cite{dkl}.

The second order contribution of the quadratic terms  in the $\delta N$ formula
gives the  one-loop contribution to the  bispectrum. 
It could be significant or even dominant.  It has been calculated
for the scalar case in Ref. \cite{bl}, and has been investigated for the 
case  of multifield inflation in for instance  Refs. \cite{cogollo,rvt}.
 The one-loop contribution from a  vector perturbation will be given
in a separate publication \cite{cesary}.

\section{Scalar field perturbation from the vacuum fluctuation}

\label{sscalar}

During inflation, both 
scalar and vector field perturbations can be generated from the
vacuum fluctuation. We begin by describing carefully the scalar field
calculation, emphasising some points that will be important when we come
to the vector field. 

\subsection{General considerations}

We shall focus on the simplest setup.
Only the few $e$-folds either side of horizon exit are considered.
Unperturbed inflation is supposed to be isotropic, and
almost exponential so that the Hubble parameter
can be taken to be  constant. 
It is assumed  that the field perturbations can be treated as 
  free fields,  so that they satisfy uncoupled 
linear field equations. Also, the 
 scalar fields are taken to live in unperturbed spacetime, which means that
the  back-reaction of the fields on the metric is ignored.
  By virtue of these features, the
 scalar field perturbations are gaussian and statistically independent, and the 
object of the calculation is to calculate their spectra.

For the calculation itself  we 
 do not need to invoke a  theory of gravity or a 
 model of inflation.  But these things are needed if one wishes to check
that the back-reaction is negligible and the field is practically free.
Assuming  Einstein gravity and slow-roll inflation,  
the check has been done as follows.
First, the  modification of the linear evolution equation to  include back-reaction
has been calculated,  both for  
 single-field  \cite{sasaki}  and multi-field  \cite{tn} inflation.
 It is  found to be
small, provided that the relevant fields
are slowly varying on the Hubble timescale, as will be the case if their potential
is flat enough for the slow-roll approximation to apply\footnote{From the form of the back-reaction, one expects this to be the case even 
for non-Einstein gravity \cite{abook}.}.
Second, the  treatment of the perturbation has 
been carried out to 
 second order \cite{maldacena,sl,sml} and third order \cite{sls,ssv}
(including the back-reaction). From this the 3-point and connected 4-point
correlators of $\zeta$ were calculated.
They were found to be negligible in accordance with the linearity
assumption.

These calculations invoke only scalar fields, which is consistent with the
assumption of isotropic unperturbed inflation. In the present paper we are going
to suppose that one or more vector fields exist during inflation. As
we noticed in Section \ref{stp}, the unperturbed (spatially homogeneous) part
of a vector field will at some level cause anisotropic unperturbed expansion.
 This  will  break the rotational invariance
of the evolution equations for the scalar  \cite{acw,kksy,watanabe,hcp,marco2},
 causing their spectra  to be
anisotropic. At the moment it is not understood how to calculate the spectra
of  scalar field perturbations in such a case, because the linear evolution
equations have singlular solutions \cite{hcp,marco2}. The generation of vector
field perturbations will also be affected by anisotropic unperturbed expansion,
though that has yet to be investigated.
As we saw in Section
\ref{stp}, the level of  anisotropy in the expansion
is expected to be small and in this
 paper we simply ignore it. 
 
\subsection{Quantum field theory}

There is no need to assume Einstein gravity during inflation. We need only the
effective action for the scalar field, valid while relevant scales are leaving
the horizon. We describe the standard scenario, in which $\phi$ is canonically
normalized. Although the calculation works for a more general potential, it 
will be  enough here to consider the quadratic case:
\be
V(\phi) = \frac12 m^2\phi^2 + \cdots
\label{quadratic} \,. \ee
The action is then 
\bea
S &=& \frac12 \int d\tau d^3x \sqrt{-g} \[  \call_\phi(\tau,\bfx) + \ldots \]  \\
\call_\phi  &=&  - \pa_\mu \phi \pa^\mu\phi
- m^2\phi^2   
\,. \label{phiaction} \eea
This is supposed to hold to good accuracy while scales of interest leave the 
horizon, with $m^2$ practically constant during that era.  
The dots indicate contributions, which generate 
 inflation if that is not already done by $\phi$ 
\footnote{If it is done by $\phi$  there is slow-roll inflation with  $\phi$ 
 the inflaton, but we are not assuming slow-roll inflation and 
still less that $\phi$ is the inflaton within that paradigm.}.

As the  field is supposed to live in unperturbed spacetime described by 
 the line element in Eq. \eqreff{rw} we can write 
\be
S = \frac12 \int  d\tau d^3x   a^2(\tau) 
%\[  (\phi')^2 -  \pa_i \phi\pa_i\phi
\[-\pa_\mu \phi \pa^\mu \phi - \frac12 a^2(\tau) m^2  - \ldots \]
, \ee
with the index $\mu$ now raised by $\eta^\mn$ instead of $g^\mn$.

{}The unperturbed field equation is 
\be
\ddot\phi + 3H\dot\phi + m^2 \phi =0
\label{scaleq} \,,\ee
where an  overdot denotes $d/dt$. 
We take inflation to be 
 practically exponential so that $a\propto \exp(Ht)$.
We assume that $\phi$ is a light field, defined as one with 
\be 
|m^2|\ll H^2 
\label{realcon}\,. \ee
If this inequality is well satisfied there will be a slow roll solution
$\dot\phi\simeq -m^2\phi/3H$, which is expected to hold more or less independently
of any initial condition. Then the fractional change in $\phi$ over one Hubble
time is much less than 1. If the inequality is only marginally satisfied
it will be of order 1.

For the first order perturbation we work with 
%conformal time $\tau=-1/aH$, and define
$\varphi\equiv a \delta\phi$. It satisfies
\be
%\[ \pa_\tau^2 + (am)^2 - 2(aH)^2 + k^2 \] \varphi(\bfk,\tau) =0 \,.
\varphi''(\bfk,\tau) + \(k^2+ a^2\tilde m^2 \)\varphi(\bfk,\tau)=0,
\qquad \tilde m^2 \equiv m^2 - 2H^2
, \label{classicaleq} \ee
where a prime denotes $d/d\tau$.
To arrive at the quantum theory we need the action for $\delta\phi$, obtained
from  \eq{phiaction}. After dropping a total derivative it is
\bea
S_{\delta\phi} &=& \frac12 \int d\tau d^3x
  \(  {\varphi'}^2 +  \pa_i \varphi\pa_i\varphi
- \frac12 a^2 \tilde m^2 \)  \\ 
 &=&  \frac12 \int d\tau d^3k
 \[    {\varphi'}^2(\bfk,\tau)  -  
\( k^2 + a^2 \tilde m^2\) \varphi(\bfk,\tau)  \] 
. \label{quantaction} \eea
For each $\bfk$ this is the action of an oscillator with time-dependent
frequency. 

We adopt the Heisenberg picture whereby the
state vector is time independent.  Promoting 
 $\varphi$ to an operator $\hat\varphi$ we  write
\be
\hat \varphi(\bfx,\tau) = \int \frac{d^3k}{(2\pi)^3} 
\[ \hat a(\bfk) \varphi(k,\tau) e^{i\bfk\cdot\bfx} + 
\hat a^\dagger(\bfk) \varphi^*(k,\tau) e^{-i\bfk\cdot\bfx} \]
\,. \label{opphieq} \ee 
The mode functions  $\varphi(k,\tau)$ satisfy the same evolution
equations as the classical perturbations $\varphi(\bfk,\tau)$. 
The former are independent of the direction of $\bfk$ because  the 
evolution equations do not pick out a preferred direction, and neither does
the initial condition that we come to shortly.

The consistent quantization of this system requires the commutation relation
\be
[ \hat a(\bfk), \hat{a}^\dagger(\bfk') ] = (2\pi)^3 \delta(\bfk-\bfk') \,,
\ee
and the Wronskian 
\be
 \varphi^*(k,\tau) \pa_\tau\varphi(k,\tau) -
\varphi(k,\tau) \pa_\tau\varphi^*(k,\tau)  = - i
\,. \ee

Well before horizon exit, $\varphi$ is a linear combination of
$\exp(\pm ik\tau)$. We  make the usual choice
\be
\varphi(k,\tau) \to \frac{ e^{-ik\tau} }{ \sqrt{2k} } \,, \label{scalarinitial}
\ee
which will be justified shortly.
We postulate a unique vacuum state,  annihilated by the $\hat a(\bfk)$,
and  take the Hilbert space to be Fock space, whose basis is built by acting on the
vacuum by products of the  creation operators $\hat a^\dagger(\bfk)$. 
The basis vectors are eigenvectors of the occupation number operator
$\hat n_\bfk = L\mthree \hat a^\dagger_\bfk  \hat a_\bfk$,
which gives the number of particles with momentum $\bfk$. The particle interpretation
can be  justified  using \eq{tmndef} for the energy momentum tensor.
It shows that  the vacuum state, with zero occupation number,
 has  momentum density and pressure  $\rho\sub{vac}=-P\sub{vac}=\Lambda^4/16\pi^2$
where $\Lambda$ is the ultra-violet cutoff. This  is set equal to zero by
 absorbing it into the scalar field potential. Then
 the  energy-momentum tensor of a generic  basis state
 is that of a gas of particles
with the relevant occupation numbers. If the occupation numbers
depend only on the direction
of $\bfk$, the momentum density and anisotropic stress vanish, leaving pressure and
energy density $P=\rho/3$.

The final step is to  assume that the time-independent state vector is close to the 
vacuum state. In other words, we assume that 
the the  occupation number $n_k$
of the quantum states (averaged over a cell of $\bfk$ space)
is much less than 1. With Einstein gravity, that  assumption is 
 mandatory if there have  been  $\Delta N\gg \ln(\mpl/H_*)$ $e$-folds of inflation
before cosmological scales leave the horizon, because the 
positive pressure $P\sim n_k (k/a)^4$ from particles with momentum of order
 $k/a$ 
would otherwise overwhelm the negative pressure $P=-3\mpl^2 H_*^2$ that is required
for inflation  \cite{abook,treview}.  The condition $\Delta N \gg \ln(\mpl/H_*)$ is 
quite mild, and will almost certainly be satisfied for the shortest cosmological scale
if inflation takes place at the usual high scale $H_*\sim 10^{-5}\mpl$.

Instead of using the negative frequency mode function  in Eq. \eqreff{scalarinitial},
one might consider using a linear combination of positive and negative frequencies.
This  corresponds to using annihilation operators $\tilde {\hat a}_\bfk$,
related to  the original ones by a Bogoliubov transformation:
\be
 \hat a_\bfk = \alpha_k \tilde {\hat a}_\bfk + \beta_k \tilde {\hat a}^\dagger_\bfk \,,
\ee
with $|\alpha_k|^2 = 1+|\beta_k|^2$. A  Fock space vector, labelled by 
the eigenvalues of 
$\tilde {\hat n}_\bfk = \tilde {\hat a}_\bfk^\dagger \tilde {\hat a}_\bfk/L^3$, 
does  not  have well-defined $n_\bfk$ and does  not have well-defined
 energy-momentum tensor either.
 In a state  where   $\tilde  {\hat n}_\bfk$ has expectation value
$\vev{\tilde {\hat n}_\bfk}$, the expectation value of ${\hat n}_\bfk$ is
\be
\vev{{\hat n}_\bfk} = \vev{\tilde {\hat n}_\bfk} + |\beta_k|^2  \( 1+ 2 
\vev{\tilde {\hat n}_\bfk} \) \,. \label{vevn}
\ee
The expectation value of the energy-momentum tensor in this state is that of
a gas with occupation number $\vev{{\hat n}_\bfk}$. 
As in the previous paragraph, it is reasonable to require this occupation number
is   much less than 1, in order to ensure that the
positive pressure of the gas will not be significant at the beginning of inflation.
Looking at \eq{vevn}, we see that this requires $|\beta_k|\ll 1$. In words,
{\em the initial mode function cannot be much different from the negative frequency
mode function in Eq. \eqreff{scalarinitial}}\footnote{Of course, it also requires that the 
state is close to the vacuum state, corresponding to $\vev{\tilde {\hat n}_\bfk}\ll 1$.}.

This argument for the choice of the negative frequency solution relies on the fact
that it minimizes the energy density {\em and} the pressure, of the gas of particles
that will be present if any other choice is made.
 The standard argument \cite{transp} invokes only the
energy density, which by itself would not be dangerous. Indeed, one is already
discounting the  vacuum energy density $\rho\sub{vac}$,  
which is permissible because it comes with $P\sub{vac}=-\rho\sub{vac}$.

\subsection{Spectrum of the perturbation}

To calculate the spectrum of $\varphi$,  we identify the
 ensemble average in \eq{spectrum} as a vacuum expectation value, with
$\varphi$ replaced by $\hat\varphi$. Then 
\be
\frac{2\pi^2}{k^3} \calp_\varphi(k,\tau)=|\varphi(k,\tau)|^2 \,.
\ee
Apart from the 
reality condition, there is no correlation between different Fourier 
components, because there is no correlation between their vacuum fluctuations
and no coupling between their evolution equations. In other words, the 
perturbation $\varphi$ is Gaussian in the linear approximation that we are
using.

The mode function is the solution of \eq{classicaleq} with the initial 
condition \eq{scalarinitial}. For $m=0$ it is
\be
\varphi(k,\tau)= - \frac i{\sqrt{2k}} \frac{ (k\tau-i) }{k\tau}
\,. \ee 
Well after horizon exit this gives
\begin{equation}
\calp_{\delta\phi} =
\frac{\calp_\varphi}{a^2} \approx \left(\frac{H}{2\pi}\right)^2 \,.
\end{equation}

Keeping $m$, we can write \eq{classicaleq} as
\be
\[ \partial_\tau^2   - \(\nu^2-\frac14\)
 \tau\mtwo + k^2 \] \varphi(\bfk,\tau)
=0 \label{classicaleq2} \,, \ee
with\footnote{As indicated we choose the positive sign.}
\be
\nu= +\sqrt{\frac94-\(\frac mH \)^2} 
\,. \ee
This is the Bessel equation
with independent solutions $J_\nu(k\tau)$ and $J_{-\nu}(k\tau)$.
The solution satisfying the initial condition is
\be
\varphi(k,\tau)=
\sqrt{\frac{\pi}{aH}}\,
\frac{e^{i\frac{\pi}{2}(\nu-\frac12)}}{1-e^{i2\pi\nu}}\,
[J_\nu(k\tau)-e^{i\pi\nu}J_{-\nu}(k\tau)] \,.
\label{perpsolu}
\ee
Well after horizon exit this gives
\be
\varphi(k,\tau) \simeq 
e^{i\frac{\pi}{2}(\nu-\frac12)} \frac{2^\nu\Gamma(\nu)}
{2^{3/2} \Gamma(\frac32)} \frac1{\sqrt{2k}} (-k\tau)^{\frac12- \nu} \,.
\label{massivemode}
\ee
The    field becomes classical in the sense that 
$[\hat\varphi (\bfk,\tau), \partial_\tau \hat\varphi (\bfk,\tau)]$ tends
to zero \cite{ls}, provided that $\nu$ is real. This condition 
corresponds to $m^2 < \frac 94 H^2$.
We reject 
the regime  $m^2\ll - H^2$ because the spectrum is too steep to be of 
interest. In any case, 
 the calculation almost certainly becomes invalid in this regime for two reasons.
First, the unperturbed field  $\phi$ will  roll rapidly away
from the origin, making it unlikely that the neglected terms of the 
potential in Eq.
\eqreff{quadratic} remain negligible over the several Hubble times that it takes
for  relevant scales to leave the horizon.
Second,  the back-reaction of the perturbation on the 
metric will probably not be negligible. These are the considerations that require
the light field condition in Eq. \eqreff{realcon}.
Applied to negative $m^2$,
this  condition is equivalent to $\nu \gsim 1$.

Well after horizon exit \eq{massivemode} gives
\bea
\calp_{\delta\phi}
&\simeq & \frac{8\pi|\Gamma(1-\nu)|^{-2}}{(1-\cos 2\pi\nu)}
\( \frac{H}{2\pi} \)^2 \( \frac{k}{2aH} \)^{3-2\nu} \nonumber \\
&\simeq& \(\frac{H}{2\pi}\)^2 \(\frac k{aH} \)^{n\sub{scalar}-1},
\label{delphispec}  \\
n\sub{scalar}-1 &=&
 3-2\nu \simeq \frac{2m^2}{3H^2} . \label{scalartilt}
\eea
The final equality is valid for $|m^2|\ll H^2$.

Instead of taking $\phi$ to be massless, one might think that a more
accurate early-time  approximation would be obtained
by keeping the mass $m$. That is not the case though, because the classicality
condition in Eq. \eqreff{realcon} means that the effect of $m$ is no bigger than
the effect of the expansion rate $H$. Because $m$ is so small, we cannot
regard it as a mass term in a flat spacetime quantum field theory.

\subsection{Spectral tilt}

\label{sscalarspectilt}

The scale dependence given by  \eqs{delphispec}{scalartilt} 
can be understood in the following way.
Soon after horizon exit, when the classical perturbation first emerges,
its spectrum is  roughly independent of $m$, and hence $\simeq (H/2\pi)^2$.
After that, the perturbation evolves according to \eq{classicaleq}
with $k^2=0$, which gives the second factor of \eq{delphispec}.

Even more simply, we can understand the scale dependence just from the 
unperturbed equation  \eqreff{scaleq}. It is the same as
\eq{classicaleq} with $k=0$ and has two independent solutions. One
is proportional to $a^{2(\nu-1)}$ and  the other to $a^{2(-\nu-1)}$.
The second solution decays relative to the first by a factor
$a^{-4\nu}$,  and one expects that it will become
negligible soon after horizon exit\footnote{Unless 
 $\nu$ is close to 1 corresponding to $n\sub{scalar}\simeq 4$.}. Using the
first solution we again arrive at \eqs{delphispec}{scalartilt}.

The curvature perturbation is given in terms of the scalar field perturbations
by  \eq{dNsc}. Let us take the initial epoch in that equation to be after all 
cosmological scales  have left the horizon, but not too long after.
Then the estimate \eq{delphispec} should apply to each scalar field perturbation.
Supposing that a single scalar 
field perturbation dominates $\zeta$, and using  the tree-level
expression for $\calpz$, the spectral index  $n$ of $\zeta$
will obviously be  equal to the spectral index $n\sub{scalar}$ of the scalar field.
The observed spectral tilt 
value $n-1\simeq -0.04$ suggests that the light field condition in Eq.
\eqreff{realcon} is very well satisfied by the relevant field.

Since the tree-level expression for $\calpz$ treats the field perturbations
linearly, one  can instead calculate the spectral index of $\zeta$
using  the `horizon-crossing trick',
whereby the initial epoch is  instead taken to be a fixed number
of Hubble times after horizon exit for the scale $k$. 
This technique allows one to easily include a slow variation of $H$,
defined by $\epsilon_H \equiv -\dot H/H^2$.
It  reduces $n$ by
an amount $6\epsilon_H$ if $\phi$ is the inflaton and by $2\epsilon_H$
otherwise. The horizon crossing technique also allows one to write down
a formula for $n$ if several scalar fields contribute, in terms of the 
first and second derivatives of the potential at horizon exit
 \cite{abook,ss,treview}.

\section{Gauge field perturbation from a time-dependent gauge coupling}

\label{stigauge}

In this section and the next, we see how a vector field
perturbation may be generated. In this section we work with 
the following effective
action during almost-exponential inflation:
\be
S=  \int d\tau d^3x \sqrt{-g} \[ - \frac14 f^2(\tau) F_\mn F^\mn  - \ldots \]
\,, \label{gaugeaction} \ee
where $F_\mn= \pa_\mu B_\nu- \pa_\nu B_\mu$ is the field strength
with $B_\mu$ a gauge field. It can be written
\be
S=  \int d\tau d^3x  \[ - \frac14 f^2(\tau) F_\mn F^\mn  - \ldots \]
\,, \label{gaugeaction2} \ee
where now the indices are raised with $\eta^\mn$ instead of $g^\mn$.

If  $f$ is time-independent 
 it can be set equal to 1 because any constant value can be
absorbed into $B_\mu$. Otherwise, $f$  represents
a time-dependent gauge coupling.
To respect invariance under time displacement,  $f$  should be a 
function of one or more fields with no explicit time dependence. 

As with the scalar field, there is no need to assume Einstein gravity
during inflation. The  other terms in the action 
are supposed to give inflation with practically constant $H$, and to
generate $f(\tau)$ without having any other effect
on the evolution of the gauge field during inflation. 
For that  to be the case,
any scalar field coupled to $B_\mu$ must have zero value (no spontaneous
symmetry breaking) with negligible quantum fluctuation around that value.

Starting with Ref.
\cite{ratra}, this action has been widely considered for the generation of 
a primordial magnetic field, and it has recently been considered \cite{ys}
for the
generation of a vector field perturbation that can generate a  contribution
to $\zeta$. In the latter context, an extension to include a mass term
is studied in Refs. \cite{VC,dkw}.

By a choice of gauge we  set $B_0$ and $\pa_j B^j$ 
equal to zero. We assume almost exponential inflation and 
 work with the perturbation 
\be
\cala_i \equiv f  \delta B_i \equiv a \delta A_i
\,. \ee
We are absorbing  $f$ into the definition of the physical field $A_i$ 
even though it is
supposed to be varying  while cosmological scales are leaving the horizon. 
At some stage
$f$ will become time-independent making   $A_i$ indeed the physical gauge
field.

The perturbation has only transverse components,  which satisfy the field equation
\be                                                       
 \cala_\lambda''(\bfk,\tau) + \( k^2-  \frac{f''}{f} \)                   
\cala_\lambda(\bfk,\tau)  = 0                                           
\,, \label{afield} \ee         
with $\lambda=L$ or $R$. The prime denotes $d/d\tau$.    
      
The quantization is just like the scalar case \cite{my}. Each  $\cala_\lambda$
has the scalar field action in Eq. \eqreff{quantaction}, with $(a\tilde m)^2$
replaced by $-f''/f$. We write
\be                                                                      
\hat\cala_i(\bfx,\tau) =  \int \frac{d^3k}{(2\pi)^3}                        
\sum_\lambda                                                             
\[                                                                       
e_i^\lambda (\hat\bfk) \hat a_\lambda(\bfk)\,                       
\cala_\lambda(k,\tau) e^{i\bfk\cdot\bfx} +                          
e_i^{\lambda *} (\hat\bfk) \hat a_\lambda^\dag(\bfk)\,                
\cala_\lambda^*(k,\tau) e^{-i\bfk\cdot\bfx}                        
\] \label{aopexp} , \ee
with the sum going only over $\lambda=L,R$. The commutator is
\be
\[ \hat a_\lambda(\bfk), \hat{a}_{\lambda'}^\dagger(\bfk') \]              
= (2\pi)^3 \delta(\bfk-\bfk')                                        
\delta_{\lambda \lambda'}, \label{vecfour} \ee
and the Wronskan of $\cala_\lambda(k,\tau)$ is $-i$.
Well before horizon exit $f''/f$ is supposed to be negligible and one 
adopts  the 
%Over some time interval
% well  before horizon exit, short on the Hubble timescale,
% we invoke flat spacetime theory and setting  $a=1$. This gives the 
initial condition 
\be                                                                           
\cala_\lambda(k,\tau) = \frac{e^{-ik\tau}}{\sqrt{2k}} 
\label{BDperp} \,,
\ee                                                         
as well as the Fock space,  and one assumes that the state is close to
the vacuum state.  
%If $f$ has negligible variation these assumptions
%are justified by the particle interpretation. 
These assumptions can be justified in the same way as for the scalar
field case.

Following Refs. \cite{my,ys} we adopt the parameterisation $f\propto a^\alpha$. 
Then \eq{afield} has the same form as \eq{classicaleq2} for the scalar field 
perturbation:
\be                                                     
\[ \pa_\tau^2 - \( \nu^2 - \frac14 \) \tau\mtwo + k^2 \] \cala_\lambda(k,\tau)
=0         \,,  \label{afield2} \ee                     
with  $\nu=\left|\alpha+\frac12 \right|$. Well after horizon exit, it leads to a 
classical 
perturbation with  the spectrum 
\be
\calp_\lambda(k,\tau) = \frac{k^3}{2\pi^2}\frac1{a^2} | \cala_\lambda(k,\tau)|^2
\,. \label{calplambda} \ee
Using the solution of \eq{afield2} with the initial condition 
in Eq. \eqreff{BDperp}
we have
\bea 
\calp_L=\calp_R\equiv \calp_+ &\simeq&  \( \frac H{2\pi} \)^2
\( \frac k{aH} \)^{n\sub{vec}-1}                                            
\,, \label{calplam2} \\
n\sub{vec}-1 &=&  3 - 2 \left |\alpha+\frac12 \right|                        
\,. \label{nvec3} \eea
The spectrum is scale invariant if $\alpha=-2$ or $\alpha=1$ \footnote
{In Ref. \cite{ys} this is given incorrectly as $\alpha=-1$. Note that the value
$\alpha=2$, advocated in Ref. \cite{my} in the context of a primordial magnetic field,
makes the energy density rather than the field perturbation scale invariant.}.

Since $\nu$ is always real, the vacuum fluctuation always gives a             
 classical perturbation after horizon exit. 
We reject $\nu\gg 1$ (equivalent to  
 $\alpha \gg1$) because the   predicted spectrum is too steep 
to be of interest.

As was pointed out in Ref.
\cite{davids}, a classical perturbation is obtained even with the standard gauge
coupling corresponding to  $\alpha=0$.
In that  case the evolution of the                                       
mode function is not affected by horizon exit and                          
$n\sub{vec}-1 =2$.  This can be traced to the                             
fact that the action is invariant under a conformal transformation of the     
metric, which means that we can go to the flat spacetime metric.        
After horizon entry during the post-inflation era,
 classicality is lost and we recover the vacuum       
state of the late-time quantum field theory, but that is of no concern       
in the present context. Of course it prevents one using the standard
action to generate 
a primordial  magnetic field (quite apart from the fact that the spectral
index would anyway be  too big for the field to be useful).

Taking $H$ to be constant, the contribution of the vector field contribution to 
$\zeta$ has spectral index index $n\sub{vec}$.     
As in \eq{flatsteep} the vector contribution could dominate on small scales, 
and even the conformal invariant tilt $n\sub{vec}-1=2$                        
might be allowed by the bound in Eq. \eqreff{extremen} though that would need       
a rather low value $N(k\sub{max})\sim 10$.                              

\section{Vector field perturbation with coupling to $R$}

\label{smodgrav}

\subsection{The action}

As an alternative to the previous case, we now 
consider the following effective action during inflation:
\be
S = \int d\tau d^3x \sqrt{-g} \[ \frac12 m_P^2 R
-\frac14 F_{\mu\nu}F^{\mu\nu}-\frac12 \(  m^2 + \frac16 R \) B_\mu B^\mu 
-\ldots \]  \,.\label{L}
\ee
%with $F_\mn= \pa_\mu B_\nu- \pa_\nu B_\mu$, and $m_P$ being the reduced Planck mass.

The third term of this action  violates gauge invariance. As a result, one 
cannot
use gauge invariance to motivate the   particular form of the kinetic term,
and one cannot use any other internal symmetry either. The
 most general quadratic kinetic
term consistent with Lorentz invariance is \cite{acw} 
\be
\call\sub{kin} = -\beta_1 \nabla^\mu B^\nu \nabla_\mu B_\nu - \beta_2
\( \nabla_\mu B^\mu \)^2 - \beta_3 \nabla^\mu B^\nu \nabla_\nu B_\mu
\,, \ee 
with $\nabla$ being the covariant derivative. Gauge invariance requires
 $\beta_1=-\beta_3$, which is the only restriction provided by
symmetry considerations. 
The action in Eq. \eqreff{L} invokes that condition, without the
justification of gauge invariance. 

The   motivation for  the action in Eq. \eqreff{L} comes, not from
symmetry considerations but because it has two remarkable
properties. One property  concerns the perturbation  $\delta B_\mu$ that is generated
from the vacuum fluctuation. As we will show in this section, the spectrum of 
the
perturbation is scale-invariant if $m=0$, for both the transverse and 
longitudinal perturbations.
 This  calculation of the spectrum  invokes no theory of gravity. 
The other remarkable property concerns the theory of gravity and will be 
described
in Section \ref{svi} (generalizing the action to include an arbitrary number of 
vector fields).  These special 
properties perhaps suggest that the action in Eq. \eqreff{L} can emerge in a natural way,
in the context of field theory or perhaps string theory.  

Much of the literature, starting with Ref. \cite{tw}, goes further and identifies
the field $B_\mu$ in \eq{L} 
with the electromagnetic field. That  requires its couplings
to other fields (including the known  Standard Model fields)
 to be of the standard gauge-invariant form even
though there is no gauge invariance\footnote
{The form of the coupling of the photon to spin half fields is completely
determined by
renormalizability, but not the form of 
its coupling to the $W^\pm$ and Higgs fields.}.
 It seems to us to be a step too far,
when  one  can as well generate a primordial magnetic field using 
the gauge invariant action of the previous section. 

We  require the other terms of the action to  generate  inflation,
without affecting the evolution of  $B_\mu$ during inflation.
For that to be the case, any terms coupling $B_\mu$ to scalar fields should
have a negligible effect. There is no reason to
suppose that such coupling occurs through the gauge-invariant
terms of the form $-\cald_\mu \phi (\cald^\mu \phi)^*$.
But if for instance a (global or gauge) $U(1)$ symmetry acts on the phase of 
$\phi$
but not on $B_\mu$ one might have a term of the form $-|\phi|^2 B_\mu B^\mu$
and then we are requiring that the the $U(1)$ is unbroken with negligible
quantum fluctuation, just as in the gauge-invariant case except that the $U(1)$
now has nothing to do with $B_\mu$. 

\subsection{Generating the field perturbation}

As the  action in Eq. \eqreff{L} contains no time derivative for the time component
$B_0$,  this component is related to the space components $B_i$ by a constraint
equation\footnote
{For a generic  choice of the kinetic term, $B_0$ becomes an independent field.
Its perturbation is considered in Ref. \cite{lim,kh}.}.
 We take the spacetime metric to be unperturbed.

The  unperturbed field has zero time component, and the space components
of the physical field $A_i=B_i/a$ satisfy
\cite{tw,RA2} 
\be
\ddot A_i + 3H\dot A_i + m^2 A_i =0 \,.
\label{weq}  \ee
This is the same as for a scalar field with mass-squared $m^2$. 

As in the previous section, we work
 with the perturbation of the physical field,
$\cala_i\equiv a\delta A_i \equiv \delta B_i$.  We expand its operator
in the form given by Eq. \eqreff{aopexp}, 
including now the longitudinal mode since
there is no gauge invariance.

Consider first the transverse modes, $\lambda=L,R$. They  satisfy  the equation 
\cite{VC,RA2}
\be
\[ \pa_\tau^2 + a^2\tilde m^2 + k^2 \] \cala_\lambda = 0
\label{tranmodeeq}
\,, \ee
where\footnote{We used the relation $R= -12H^2$, valid during exponential 
inflation.} 
\be
\tilde m^2 = m^2+ \frac16 R=    m^2 - 2 H^2
\,. \label{masscondition} \ee
This is the same as  
 for a scalar field with mass-squared $m^2$. The action for each of
$\cala_\lambda$ is also the same \cite{marco2}.
%Invoking flat spacetime
%field theory well before horizon exit gives the initial condition in Eq.
%\eqreff{BDperp} of the previous section, and solving the mode function
%gives as in the scalar case 
We adopt the initial condition, the Fock space, and the vacuum state assumption,
with the same justification 
 as in the scalar field case. Then
\bea
\calp_+
&\simeq & \(\frac{H}{2\pi}\)^2 \(\frac k{aH} \)^{n\sub{vec}-1} \,,
\label{delaspec2}  \\
n\sub{vec}-1 &=&
 3-2\nu \simeq \frac{2m^2}{3H^2}\,,\qquad \nu\equiv \sqrt{\frac94 - 
\frac{m^2}{H^2} }
\,. \label{vectilt2}
\eea
A classical perturbation is generated if $\nu$ is real corresponding
to $m^2<9H^2/4$.
As with the scalar case, we reject the case $m^2\ll -H^2$. The spectrum is too
steep to be of interest, and anyway   the evolution of $A_i$ would be so
 rapid
that additional terms in Eq. \eqreff{L} (required to stabilize $A_i$) could 
hardly
remain negligible  over the several Hubble times that it takes for
cosmological scales to  leave  the horizon. We therefore require
\be
-H^2 \lsim m^2 < \frac 94 H^2
\,. \label{mrange} \ee
As advertised, the  tilt vanishes if $m=0$.

Now we discuss
%specify for the first time 
the quantization of the longitudinal perturbation. 
Its mode  function  satisfies \cite{VC}
\be
\[ \pa_\tau^2 + \frac{2k^2 a H}{\( k^2+a^2\tilde m^2 \)}\pa_\tau
+ \(k^2 + a^2\tilde m^2\) \] \cala\sub{long} = 0
\,. \label{longmodeeq} \ee
For $m=0$ corresponding to $\tilde m^2=-2H^2$, 
the independent solutions (given here for the first time) are
\be
\cala\sub{long}^{\pm}(k\tau) \propto
\( - k\tau + \frac2{k\tau} \pm 2i \) e^{\mp ik\tau}
\,. \label{longsol} \ee
We see that the solutions are  regular even at the point  where
 the round bracket  in \eq{longmodeeq} vanishes.

We can show that the solution of \eq{longmodeeq}  is non-singular even for
$m^{2}\neq0$. This can be done by using the Frobenius method for
differential equations with regular singular points (see for example
Ref. \cite{Rabenstein}). First we make a change of variables
\begin{equation}
y\equiv\left(\frac{k}{a\left|\tilde{m}\right|}\right)^{2}-1 \,,
\end{equation}
with $y$ varying in the region $-1<y<\infty$. Eq. (\ref{longmodeeq}) with this
transformation translates into the form
\begin{equation}
\left[\partial_{y}^{2}-\frac{1}{2}\frac{\left(y+2\right)}{y(y+1)}\partial_{y}+\frac{\left|\tilde{m}^{2}\right|}{H^{2}}\frac{y}{4\left(y+1\right)^{2}}\right]\mathcal{A}_{\mathrm{long}}=0 \,,\label{eq:EoM-y}
\end{equation}
with $\tilde{m}^{2}<0$ and the regular singular point at $y\rightarrow0$.
The general solution of this equation can be found using the ansatz
\begin{equation}
\mathcal{A}_{\mathrm{long}}=\sum_{n=0}^{\infty}D_{n}y^{s+n} \,,\label{eq:A-ansatz}
\end{equation}
where $D_{0}\neq0$. In this case the series in Eq. (\ref{eq:A-ansatz})
is convergent at least in the region $-1<y<1$ without a singular
point. We will show that it converges even at this point and that
the ansatz in Eq. (\ref{eq:A-ansatz}) gives two independent solutions. To
show this let us substitute Eq. (\ref{eq:A-ansatz}) into Eq. (\ref{eq:EoM-y})
giving
\begin{eqnarray}
\sum_{n=0}^{\infty}D_{n}\left[4\left(s+n\right)\left(s+n-2\right)y^{s+n-2}+8\left(s+n\right)\left(s+n-\frac{7}{4}\right)y^{s+n-1}+\right. \nonumber \\
\left.+4\left(s+n\right)\left(s+n-\frac{3}{2}\right)y^{s+n}+\frac{\left|\tilde{m}^{2}\right|}{H^{2}}y^{s+n+1}\right]=0 \,. \label{eq:EoM-series}
\end{eqnarray}

In order for the equality in Eq. (\ref{eq:EoM-series}) to be valid, coefficients
in front of each $y$ with the same power must vanish. The coefficient
in front of the term with the smallest power, i.e. $y^{s-2}$, is
$4D_{0}s\left(s-2\right)$. Because $D_{0}\neq0$, from the indicial
equation $s\left(s-2\right)=0$ we find
\begin{equation}
s=0 \,,\quad\mathrm{or}\quad s=2 \,.
\end{equation}

Because these two solutions differ by an integer, it might be alarming
that the general solution of Eq. (\ref{eq:EoM-y}) might involve the
logarithm. However, by closer inspection of Eq. (\ref{eq:EoM-series})
we find that the coefficient $D_{2}$ of the series with $s=0$ is
arbitrary, thus the power series in Eq. (\ref{eq:A-ansatz}) with $s=0$
and $s=2$ give two independent solutions. And because the series
does not involve negative powers of $y$, i.e. $s\geq0$, it converges
at the singular point $y\rightarrow0$.

The action corresponding to \eq{longmodeeq} is\footnote{This is given for the case $\tilde m^2=-2H^2$ in Ref.
 \cite{marco2}, and it  can be derived by perturbing the full action.
Of course it is unique only up to a total derivative.}
\bea
S\sub{long} &=& \frac12 \int d\tau d^3k \call \,, \\
\call &=&  (a\tilde m)^2
\[ \frac{ |\cala'\sub{long}(\bfk,\tau) |^2 }{k^2 + (a\tilde m)^2} - 
| \cala\sub{long}(\bfk,\tau) |^2 \] \,. \label{marcoaction}
\eea

To set the initial condition well before horizon entry we define 
%$\cala\sub{long}=(k/a|\tilde m|) \tilde\cala$.
$\tilde \cala = (a |\tilde m|/k) \cala\sub{long}$.
In the regime $a|\tilde m|\ll k$,
\be
\call = \pm \(  |\tilde \cala'|^2 - k^2 |\tilde \cala|^2 \)
, \ee
where the sign $\pm$ is that of $\tilde m^2$, hence negative for the 
case of interest  $\tilde m^2 \simeq -2H_*^2$.  

 Except for the negative sign this is 
same as for the scalar field case.  To quantize it we 
assume the same  initial condition $\tilde \cala = \exp(-ik\tau)/\sqrt{2k}$,
and adopt the  vacuum state. The justification for these assumptions is similar
to the one that holds for the scalar field (and transverse vector field),
but not identical because of the negative sign. Because of this sign,
 occupied initial states would  have negative energy density and pressure, 
$P=\rho/3 \sim -n_k (k/a)^4$. As the pressure is negative it is not dangerous
for inflation. Instead, it is the negative energy density that is dangerous. 
As the total energy
density is required to be positive, the negative contribution of 
occupied states has to be less than the total at the beginning of inflation.
Assuming as before $\Delta N\gg \mpl/H_*$ $e$-folds of inflation before 
cosmological scales leave the horizon, this again requires occupation number
much less than 1, 
justifying both the choice of initial mode function and the assumption of
the vacuum state.

The  spectrum $\calp\sub{long}$ is  given by \eq{calplambda}.
For $m=0$, corresponding to $\tilde m^2=-2H^2$,  we find well after horizon exit
\be
\calp\sub{long} = 2\( \frac{H}{2\pi} \)^2 = 2\calp_+ \,.
\ee
This corresponds to $r\sub{long} =2$, which according to the discussion at the
end of Section \ref{streespec} means that the vector field perturbation
cannot generate the dominant contribution to the curvature perturbation.

It has been suggested \cite{hcp,marco2} that the
 action in \eq{marcoaction}  does not correspond to
a well defined quantum field theory for negative $\tilde m^2$. 
 We have demonstrated
that there is a well defined quantum field theory even in this case.
Before the epoch $|a\tilde m|^2 =k^2$, a  negative  
 $\tilde m^2$ corresponds to a negative kinetic term in the action.
This will cause some degree of   instability when
more terms are included in the action,  corresponding to the interaction of
$\cala\sub{long}$ with other fields and/or gravity.
But such interactions are assumed to be negligible whenever one considers
the generation of a gaussian classical field perturbation from the vacuum
fluctuation, and as we mentioned already has been  justified
for both scalar and vector field perturbations.
 In this connection, it is important to realise that the
 the negative sign holds only before the epoch
$|a\tilde m|^2 =k^2$ which is around the time of horizon exit.
Also, that  only a limited number of $e$-folds of inflation take place
between the emergence of $k/a$ from the Planck scale and horizon exit,
which means that there is only a limited amount of time for the presumably
small interactions  of $\cala\sub{long}$ to have any effect.
After horizon exit, the evolution at each location is given by
the classical expression in Eq. \eqreff{weq} and we have no more need of the 
quantum theory. According to the classical expression $A_i$
is slowly varying. It moves  towards zero if $m^2$ is positive.
If instead $m^2$ is negative   moves towards  the vev of $A_i$.
That vev  will be at the minimum of the   potential $V(B_\mu B^\mu)$,
whose leading term $m^2 B_\mu B^\mu/2$ is displayed in the action
in Eq. \eqreff{L}.

\section{Vector curvaton}

\label{svc}

We have described two mechanisms that can generate a vector field perturbation
from the vacuum fluctuation. In this section and the next we describe two
mechanisms by which such a perturbation can give a contribution to the curvature
perturbation. We begin in this section with the vector curvaton mechanism
\cite{VC}. This is the curvaton mechanism
 \cite{curvaton1,curvaton2,curvaton3,luw}, using a vector field instead of the usual
scalar field.

The vector curvaton field $A_i(\bfx,\tau)$
 is  smoothed on a scale somewhat below
the shortest cosmological scale and it has a perturbation $a^{-1} \cala_i=\delta A_i$. 
After horizon exit during inflation, the spatial gradient of 
$A_i$ becomes negligible and it evolves at each point as an unperturbed field.
In the simplest curvaton scenario, which we adopt, the evolution is negligible
during and after inflation, until some epoch when $A_i$ begins to oscillate.
 At this epoch, there is supposed to be
Einstein gravity and the effective action  is supposed to be
\be
S = \int d\tau d^3x \sqrt{-g} \[ 
-\frac14 F_{\mu\nu}F^{\mu\nu}-\frac12  m^2  B_\mu B^\mu 
-\ldots \]  \,. \label{L4} \ee
This is the action of a massive vector field, living in the expanding Universe
which is taken to be unperturbed. When $H$ falls below the mass $m$,
the field begins to oscillate with angular frequency $m$. As the spatial gradient
is negligible, the oscillation is a standing wave whose initial amplitude
 varies with position.

As originally proposed, the 
vector curvaton scenario   generates the perturbation
$\delta A_i$ with  essentially the action in Eq. \eqreff{L}, 
 taking  $m^2$ to be a constant
parameter  which during inflation is negligible.  For the present purpose
there is no need to say how the perturbation is generated.

The energy density of the oscillation is, in terms of the physical field
$A_i=B_i/a$, 
\bea
\rho_A(\bfx,t)  &\simeq &\frac12 m^2 | A(\bfx,t)|^2  
\(\frac{a\sub{start}}{a(t)}\)^3 \\
&=&  \frac12 m^2 \( | A|^2 + 2 A_i\delta { A}_i(\bfx) + 
\delta { A}_i(\bfx) \delta { A}_i(\bfx)
 \)  \(\frac{a\sub{start}}{a(t)}\)^3
\,, \label{oscillationamp} \eea
where $a\sub{start} $ is the scale factor just before the oscillation starts.
In the second line,  $ \bfA$ is the unperturbed value just before the oscillation
starts and $\delta \bfA(\bfx)$ is its perturbation. 
The oscillation amplitude falls like $a\mthreehalf$, 
and is practically constant 
during one oscillation. As a result,
 the stress is practically zero just as in the scalar
field case \cite{VC}. 
We take the decay to be instantaneous, which from the scalar field case
we know will be an adequate approximation.

The contribution of $\rho_A$ to the total energy density is supposed to be
initially negligible, and with it the contribution $\zeta_A$
of $\delta A_i$ 
to $\zeta$. But the oscillation is supposed to take place in a 
radiation background, so that $\rho_A/\rho$ grows like $a(t)$ and
$\zeta_A$ becomes significant.

To calculate  $\zeta_A$ we will use the
 following  expression \cite{luw}: 
\bea
\zeta_A &=&\frac13 \Omega_A \frac{\delta \rho_A}{\rho_A} \,, \\
\Omega_A &\equiv &  \frac{3\rho_A}{ 3\rho_A + 4 \rho_r} \simeq \frac{\rho_A}{\rho}
\,, \label{zetaexp} \eea
where $\rho=\rho_A+\rho_r$.
  This   expression is valid to first order
in  $\delta\rho_A$, which  is evaluated
 on a  `flat' slice where  $a(\bfx,t)$ is unperturbed.

We take the curvaton to decay instantly (sudden-decay approximation)
and evaluate $\zeta_A$  just before the curvaton decays, assuming that
$\zeta$ is constant thereafer.
 The final equality in \eq{zetaexp} 
is justified because the sudden decay approximation gives an error of similar
magnitude, both errors disappearing  in the limit $\Omega_A=1$.
Evaluating $\delta \rho_A$ to  first order we have
\be
\zeta_A = \frac23  \Omega_A  \frac{A_i \delta A_i}{|A|^2}
\,. \label{zetaafirst} \ee

%Since $A_i$ has nonzero mass, we expect that the longitudinal fraction
%$r\sub{long}$ will be nonzero. As  argued in Section \ref{smodgrav}, we expect
%$r\sub{long}$ to be roughly of order 1 if the perturbation $\delta A_i$
%is generated  during inflation with the action as in Eq. \eqreff{L}.
The tree-level contribution to the spectrum is
\be
\calp_{\zeta_A}(k)  = \frac49 \frac{\Omega_A^2}{|A|^2} \calp_+(k)
 \[ 1 + \( r\sub{long}-1\) 
(\hat{\bf A}\cdot\hat{\bf k}
)^2 \]
\label{calpza} \,,
 \ee
where \mbox{$\hat \bfA\equiv \bfA/|A|$}. 

The spectum $\calp_+(k)$ is to be evaluated just before the oscillation starts.
In Ref. \cite{VI} it is taken to be the same as that at the initial epoch during
inflation and that in turn is supposed to be generated from the action
in Eq. \eqreff{L}. Then  $\calp_+$ is given by \eq{delaspec2} with $n\sub{vec}$
practically equal to 1.

Evaluating $\delta\rho_A$ to  second order we have \cite{DY}
\be
\zeta_A = \frac23 \Omega_A \frac{A_i \delta A_i}{|A|^2} + 
\frac13 \Omega_A \frac{\delta A_i \delta A_i}{|A|^2}
\,. \label{zetasecond} \ee
This is valid only for $\Omega_A\ll 1$. To handle the case $\Omega_A\simeq 1 $
one could go to second   order in $\delta\rho_A$, or much more
simply evaluate $N$ and hence $\delta N$ directly\footnote{To first order in $\delta\rho_A$, one finds by that method
$N_A^i= 2\Omega_A A_i/3|A|^2$, in agreement with \eqs{zetaafirst}{zetasecond}.}.
All of this is the same as for a scalar field contribution, where
the evaluation of $N$ was done in Ref. \cite{DY}. We shall not pursue the
case $\Omega_A\simeq  1$ in the present paper.

Our \eq{zetaafirst} is  Eq.~(64) of Ref. \cite{VC}, 
generalized to allow $\Omega_A< 1$ and written 
to exhibit manifest invariance under rotations. The spectrum $\calp_{\zeta_A}$
was not calculated in Ref. \cite{VC} but it was implicitely assumed to be
rotationally invariant so that it could be the dominant contribution.

In accordance with the discussion at the end of Section \ref{streespec},
this realisation of the   vector curvaton mechanism 
%with a single vector field 
cannot give the  dominant contribution to $\zeta$. 
%if $(r\sub{long}-1)$ happens to be less than $30\%$ or so. 
It could do so by invoking   several  vector curvaton fields. 
We note that the case of  several  scalar curvaton fields
has   been considered in Ref. \cite{nvaton}. 

\section{Vector inflation}

\label{svi}

Recently, it has been proposed \cite{VI} (see also Refs. \cite{koivisto,gmv2,gvnm})
  that 
inflation can be driven by
a large number of independent vector fields.  They considered only
the unperturbed case, and invoked the large number to make the unperturbed
metric practically isotropic. We   consider the perturbation. 

The action is  \eq{L}, extended to include many vector fields:
\be
S = \int d\tau d^3x \sqrt{-g} \left\{ \frac12 m_P^2 R -\sum_b
\[ \frac14  F^{(b)}_{\mu\nu}F^{(b)\mu\nu}-
\frac12 \(  m^2 + \frac16 R \) B^{(b)}_\mu B^{(b)\mu } \]
-\ldots \right\}  \,.\label{L2} \ee
As it is supposed to apply throughout inflation (starting with the approach
of  horizon exit for the largest cosmological scale $k\sim H_0$), the
 additional terms are supposed to be negligible 
throughout that era, and not just while
cosmological scales are leaving the horizon. Also, the
 action is supposed to define the theory of gravity as well as the
dynamics of the vector fields. 

Consider first the unperturbed fields $B^{(b)}_i(\tau)$.  Because each of them
has a direction, the expansion  is not generally isotropic but the anisotropy
can be negligible if there is a large number of 
randomly oriented fields \cite{VI} which is assumed. Given a large number of fields,
the randomness assumption is well justified because, as stated 
 in Section \ref{sdeln},  the unperturbed field values are defined as spatial
averages within a chosen box, whose location is random. By the same token,
it does not seem  reasonable to replace the randomness assumption by the assumption
that there are three  fields whose unperturbed values are orthonormal, though that
would also give unperturbed spacetime \cite{triplet,armendariz}\footnote{The choice might be 
justified on anthropic grounds if isotropic expansion was favoured on those
grounds but that there is no suggestion that  such is  the case. In particular
there  is no suggestion that
the  $30\%$ or so of anisotropy  allowed by present data is anthropically 
disfavoured.}.

Varying the action with respect to an unperturbed field, one finds
that  \eq{weq} is satisfied. Varying the action instead with 
respect to the spacetime metric gives   the right hand side of the
Einstein field equation, which we take as the definition of the energy momentum
tensor. For a generic  spacetime, the term coupling $R$ to the vector fields
would make the form of this energy momentum tensor dependent on the metric;
in other words it would modify Einstein gravity. Remarkably though, the modification
is negligible when spacetime is practically unperturbed \cite{VI}. As a result
we have the usual expressions, depending only on the vector field:
\bea
\rho &=& \frac12 \sum_{b,i} \[ \( \dot  A_i^{(b)} \)^2 + m^2  \( A_i^{(b)} \)^2 \] \,,
\label{rho} \\
P &=& \frac12 \sum_{b,i} \[ \( \dot  A_i^{(b)} \)^2 - m^2  \( A_i^{(b)} \)^2 \]
\,. \label{p} \eea
The Friedmann equation therefore  takes the usual form, $3 m_P^2H^2 = \rho$.

From \eqss{weq}{rho}{p} we see that  each component of the unperturbed field
is equivalent to a scalar field. In the regime $H^2\gsim m^2$ there is inflation,
with 
\be
H^2\simeq\frac16\frac{m^2}{m_P^2}\sum_b |\bfA^{(b)}|^2  \label{vecfr} \,.
\ee
It follows that the number of $e$-folds to the end of inflation is given
by the same expression as in the scalar field case \cite{starobinsky,al}:
\be
N\simeq \frac{1}{4 m_P^2}\sum_b|\bfA^{(b)}|^2 \,.
\label{nvinf} \ee

Now we consider 
%for the first time 
the curvature perturbation generated by
vector inflation. It turns out to be practically the same as if the field components
are replaced by scalar fields and that case has already been worked out using the
$\delta N$ formalism \cite{al}. 
 The derivatives of $N$ for use in the $\delta N$ formula are
given by  \eq{nvinf}:
\be
N_{A^{(b)}}^i=\frac{A_i^{(b)}}{2m_P^2}\,, \qquad 
N_{A^{(a)}A^{(b)}}^{ij} =\frac 1{2m_P^2}\delta_{ij}\delta_{ab}\,.
\ee
The
transverse spectrum $\calp_+$ of the field perturbations 
are given by \eqs{calplam2}{nvec3} (the same as for a scalar field)
 and the longitudinal spectra are $\calp\sub{long}= 2 \calp_+$.
%where $r\sub{long}$ is an uknown constant expected to be roughly of order 1.

The spectrum $\calpz$ is given by \eq{Pzeta2} (without the scalar
contribution), summed over all of the
vector fields using $\bfn_{A^{(b)}} = \bfA^{(b)}/2 m_P^2$.
Since $m^2\ll H^2$, we have  $\calp_+\simeq (H_k/2\pi)^2$ for each field,
where $H_k$ is the Hubble parameter when the scale $k$ leaves the horizon.
 Since there are a large
number of randomly oriented fields we can pretend that they all have the
same magnitude when evaluating the second term. Since the average of
$\cos^2$ is $1/2$, this gives
\be
\calp_\zeta(k)  =  
%N \( \frac{1+r\sub{long}}2 \)
\frac32 N
\( \frac{ H_k}{2\pi m_P} \)^2 
\,. \ee
Except for the factor $3/2$, the spectrum is the same as was found for
 the scalar field case \cite{al}. Such a result is independent of the number of fields.  

Assuming that $N\simeq 55$ $e$-folds of inflation take place after
the observable Universe leaves the horizon, the observed magnitude
 of  $\calpz$ is reproduced if $H \simeq 10^{14}\GeV$ at the end of 
inflation\footnote
{With a standard cosmology after inflation,
this  high inflation scale indeed corresponds to  $N\simeq 55$.}. 
The non-gaussianity is negligible and the 
 spectral index is $n=1-2/N$. 

In Ref. \cite{gmv2} (see also Ref. \cite{gvnm})  the  tensor perturbation $\delta h_{ij}$
is also considered, actually for  a wide  class of vector inflation models including  the 
one considered here. The tensor perturbation  is supposed to live in unperturbed
spacetime, as in the standard calculation described in Section \ref{stp}. But because
the action in Eq. \eqreff{L2} does  not correspond to Einstein gravity its  linear
evolution  equation differs from \eq{hijeqn}. It is found that
$\delta h_{ij}$ can have significant time dependence,  which makes it difficult to see how
the prediction can be compared with observation.

To avoid these problems, one should go to the  Einstein frame by making a conformal
transformation of the metric. Suppose that we have an  action  of the form
\be
S = \int d\tau d^3x \sqrt{-g} \[ f R + \call\sub{matter}(g_\mn,\cdots) \]
, \ee
with $f$ being any scalar function of bosonic fields, and $\call\sub{matter}$
a function of the bosonic fields that is obtained from a flat spacetime expression
through the replacements $\eta_\mn\to g_\mn$ and $\pa_\mu\to \bigtriangledown_\mu$ where
$\bigtriangledown_\mu$ is the covariant derivative\footnote
{This means that $\call\sub{matter}$ is obtained
from a  flat spacetime expression using the equivalence principle even though 
the dependence of $f$ upon the bosonic fieds means that the full lagrangian violates
the equivalence principle. Using the vierbein formalism,  $\call\sub{matter}$
can include spinor fields.}.  Now we make   a conformal transformation of the metric,
 $\tilde
g_\mn =\exp(2\Omega)  g_\mn$ with $\exp(2\Omega) \equiv 2f/m_P^2$.
 After dropping a total derivative this gives \cite{ok}  
\bea
\tilde S &=& \int d\tau d^3x \sqrt{-\tilde g} \( \frac12 m_P^2  \tilde R + 
\call\sub{matter} \) \\
\tilde \call\sub{matter}  &\equiv& 
2 \tilde \bigtriangledown_\mu \Omega \tilde \bigtriangledown^\mu \Omega  + e^{2\Omega} \call\sub{matter}
( g_\mn(\tilde g_\mn),\cdots) \
. \eea
 The usual application \cite{msf,ok}
is to slow roll inflation, with $f(\phi)$ a function
of just the inflaton field $\phi$.  Then the conformal transformation 
%changes the form of the inflaton potential, and it 
%multiplies the kinetic term of  the inflaton field by  $m_P^2/2f(\phi)$. 
just gives $\phi$ a non-canonical kinetic term.
The single field $\phi$
can be redefined to have a canonical kinetic term, so that we again
have slow roll inflation though with a different potential.

In our case, 
\be
2f= m_P^2 + \frac16 \sum_b B_\mu^{(b)} B^{(b)\mu}
.\ee
Since  $f$ is slowly varying, there is almost 
exponential
inflation in the Einstein frame just as in the original frame, with practically
the same Hubble parameter $H_*$. Now though, 
$\tilde\call\sub{matter}$ is a complicated function,  making the
Einstein frame  completely unsuitable for the calculation of
the vector field perturbations, and hence of the curvature perturbation. 
It is however the one in which one should calculate
the tensor perturbation. 

Indeed,  the standard first-order cosmological perturbation
theory calculation described in Section \ref{stp} will apply in the Einstein frame,
provided that the number of vector fields is large enough to make the 
anisotropic stress tensor negligible.
The linear evolution of the tensor perturbation in the Einstein frame is  then given
by  \eq{hijeqn}, and the spectrum is given by \eq{calpten}. 
The tensor fraction 
$r\equiv \calp\sub{ten}/\calpz$ is therefore given by the same formula
as in the scalar field case, which is  \cite{al} $r=8/N$.

Unfortunately these  combined predictions  for $n$ and $r$ are 
disfavoured by observation \cite{wmap5}. Making the masses unequal would
make the spectral index even less than one without altering $r$ \cite{al},
which increases the disagreement with observation. Therefore, the 
dominant contribution to $\zeta$ probably has to be generated after inflation.

Finally, we mention that in Ref.
\cite{gmv2}, more  general vector inflation models are constructed, with 
the mass term replaced  by a more general potential.
These  models are  again equivalent to  models with a large number of
scalar fields. The spectra of the field perturbations are the same as before
(since they invoke only almost exponential inflation without specifying its
origin) but their effect on $\zeta$ depends in general on what happens at the
end of inflation \cite{myend}, 
which is determined by other terms in the action.

\section{Conclusions} 

\label{scon}

Until recently, it has been assumed that only scalar fields play a significant
role during inflation.  
Then the spectrum of the curvature
perturbation is statistically isotropic and homogeneous, and so are higher 
correlators that would correspond to non-gaussianity. Now, it is being 
recognised that vector fields might be significant during inflation.
In that case, the correlators of the curvature perturbation will at some
level be anisotropic (though still homogeneous). The anisotropy will
occur if an unperturbed vector field causes anisotropy in the expansion
rate, because that will cause the correlators of the scalar field perturbations
to be anisotropic. It will also occur if a vector field perturbation 
contributes significantly to the curvature perturbation. 

In this paper we have for the first time 
given expressions for the spectrum and bispectrum
of the curvature perturbation, which include the second of these 
effects for a generic vector field. 

On the theoretical side, we have for the first time considered the generation
from the vacuum of a longitudinal vector field component, which will be present
in the absence of gauge invariance. Taking its action to be that in \eq{L},
we have shown that it can be described by a quantum field theory,
according to which its spectrum is twice that of the transverse field
components. 

%other which postulated a reasonable sho Within the e have shown that 
%find that field theory provides no  well-motivated 
% initial condition for the mode function, which defines the vacuum state of
%the theory. Accordingly, the predictions contain a parameter $r\sub{long}$,
%the ratio of the longitudinal to the transverse spectrum, that cannot be
%calculated within the context of field theory. However, 
%on the reasonable
%assumption that the mass-squared  $m^2$ 
%of the vector field becomes irrelevant in the regime of interest 
%$|m^2|\lsim H^2$, we expect $r\sub{long}$ roughly of order 1.

We have also given  general formulas for  the statistical anisotropy
of the spectrum and bispectrum, in terms of the longitudinal and transverse
spectra of the nearly-gaussian vector fields.
On the observational side, this leads to a very interesting situation
 regarding statistical anisotropy, which is very similar to that 
obtained a 
few years ago regarding non-gaussianity. The accepted mechanism for generating
$\zeta$, from the perturbation of the field(s) responsible for slow roll inflation,
predicted  negligible non-gaussianity
\cite{maldacena}, and gaussianity was taken for granted in most
early analysis of the observations. 
 Starting with the curvaton model \cite{curvaton1,curvaton2,curvaton3}
it was found \cite{luw} that instead the non-gaussianity could be
large,   and this motivated an intensive search for non-gaussianity.

Now that vector field contributions to the curvature perturbation are
under consideration, statistical isotropy, which previously was taken
for granted, should be reconsidered. We look forward to the opening up of
a new area of research, in which predictions for the anisotropy are
developed, and confronted with observation. In this context it should
be emphasised that the bispectrum (and higher correlators) of the
curvature perturbation might be completely anisotropic\footnote{See for instance Ref. \cite{dkl}.}, corresponding
to the dominance by one or a few vector fields. 

%%%%%%%%%%%%%%%%%%%%%%%%%%%%%%%%%%%%%%%%%%%%%%%%%%%%%%%%%%%%%%%%%%%%%%
\section{Acknowledgments}
%%%%%%%%%%%%%%%%%%%%%%%%%%%%%%%%%%%%%%%%%%%%%%%%%%%%%%%%%%%%%%%%%%%%%%
D.H.L. thanks Sean Carroll, Adrienne Erickcek, Marc Kamionkowski,
Shuichiro Yokoyama, and Jun'ichi Yokoyama  for useful correspondence. Y.R.
and D.H.L. thank  C\'esar Valenzuela-Toledo for useful comments and 
discussions.
K.D. and D.H.L. are  supported by PPARC grant PP/D000394/1 and
by EU grants MRTN-CT-2004-503369 and MRTN-CT-2006-035863.
M.K. is supported by the Lancaster University Physics Department.
 Y.R. is supported by COLCIENCIAS grant No. 1102-333-18674 CT-174-2006, 
DIEF (UIS) grant No. 5134, and the ECOS-NORD Programme grant No. C07P02.

%%%%%%%%%%%%%%%%%%%%%%%%%%%%%%%%%%%%%%%%%%%%%%%%%%%%%%%%%%%%%%%%%%%%%%

%%%%%%%%%%%%%%%%%%%%%%%%%%%%%%%%%%%%%%%%%%%%%%%%%%%%%%%%%%%%%%%%%%%%%%

\end{document}